\documentclass{revtex4}
\usepackage{amsthm,amssymb,amsmath}				
\usepackage{graphicx,comment}
\usepackage{float}   
 \usepackage{xcolor}

\begin{document}

\title{Effect of nonlocal transformations on the linearizability and exact solvability of the nonlinear generalized modified Emden type equations}
\author{Omar Mustafa}
\email{omar.mustafa@emu.edu.tr}
\affiliation{Department of Physics, Eastern Mediterranean University, G. Magusa, north
Cyprus, Mersin 10 - Turkey.}

\begin{abstract}
\textbf{Abstract:}\ The nonlinear generalized modified Emden type equations (GMEE) are known
to be linearizable into simple harmonic oscillator (HO) or damped harmonic oscillators (DHO) via some nonlocal transformations. Hereby, we show that the structure of the nonlocal transformation and the linearizability into HO or DHO determine the nature/structure of the dynamical forces involved (hence, determine the structure of the dynamical equation). Yet, a reverse engineering strategy is used so that the exact solutions of the emerging GMEE are nonlocally transformed to find the exact solutions of the HO and DHO dynamical equations. Consequently, whilst the exact solution for the HO remains a textbook one, the exact solution for the DHO (never reported elsewhere, to the best of our knowledge) turns out to be manifestly the most explicit and general solution that offers consistency and comprehensive coverage for the associated under-damping, critical-damping, and over-damping cases (i.e., no complex settings for the coordinates and/or the velocities are eminent/feasible).  Moreover, for all emerging dynamical system, we report illustrative figures for each solution as well as the corresponding phase-space trajectories as they evolve in time.


\end{abstract}

\maketitle

\section{Introduction}

Nonlinear oscillators are commonly used to describe a variety of physical systems ranging from elementary pedagogical implementations to more advanced physical phenomena like atmospheric, plasma, nonlinear optics, electronics,
biophysics, etc (c.f., e.g., \cite{Chandrasekar-PRE 2005} and related references cited therein). They represent more realistic models than the simple harmonic oscillator (HO). The so called generalized modified Emden
equation (GMEE)%
\begin{equation}
\ddot{q}+\left( 3k\,q+a_{_{1}}\right) \,\dot{q}%
+a_{_{2}}q+a_{_{1}}kq^{2}+k^{2}q^{3}=0  \label{GMEE0}
\end{equation}%
is one of such nonlinear oscillators, where $a_{_{1}},a_{_{2}},k\in 
\mathbb{R}
$ \cite{Chandrasekar 2006, Chandrasekar 2012}. It collapses into the modified Emden equation MEE for $a_{_{1}}=0$, into damped harmonic oscillator (DHO) for $k=0$, and into simple harmonic oscillator (HO) for $a_{_{1}}=k=0$.  Its integrability properties are discussed in details (e.g.,  \cite{Chandrasekar-PRE 2005,Chandrasekar 2006,Chandrasekar 2012,Leach 1985,Leach 1993,Chandrasekar1 2006,Chandrasekar 2007}) and lie far beyond the scope of the current proposal. Such a nonlinear equation finds
its feasible applications in, for example, the equilibrium configuration of spherical gas clouds \cite{Leach 1985}, the spherically symmetric relativistically gravitating mass \cite{Mc Vittie 1980}, the Yang-Mills boson gauge theory \cite{Yang-Mills 1954}, etc. The linearizability of such a second-order ordinary differential equation (ODE)
into either linear HO or linear DHO is made possible through the interesting nonlocal transformation%
\begin{equation}
U\left( t\right) =q\left( t\right) exp\left({\int f\left(q( t)\right) 
dt}\right)\Longleftrightarrow \frac{\dot{U}\left( t\right) }{U(t)}=\frac{\dot{q}(t)%
}{q(t)}+f\left(  q(t)\right)    \label{NLT}
\end{equation}%
by Chandrasekar et al. \cite{Chandrasekar 2006}. Obviously, not only the structure of $f\left( q(t)\right) \in 
\mathbb{R}
$ in (\ref{NLT}) but also the linearization process into either linear HO or DHO would determine the nature/structure of the dynamical force involved in the problem at hand. Chandrasekar et al. \cite{Chandrasekar 2006} have considered a variety of $f\left( q\left( t\right) \right) $ structures and worked out the corresponding integrability properties and solutions. 

In the current methodical proposal, nevertheless, we shall use a linear form for $f\left( q\left( t\right) \right) $ and consider
\begin{equation}
f\left( q\left(
t\right) \right) =\alpha q_{_{\pm }}\left(
t\right) \pm \zeta,
\label{f(q)}
\end{equation}%
to imply that the transformation (\ref{NLT}) now reads 
\begin{equation}
U_{_{\pm }}\left( t\right) =q_{_{\pm }}\left( t\right) \exp \left( \int
\left( \alpha q_{_{\pm }}\left( t\right) \pm \zeta \right) dt\right)
\;\Longleftrightarrow \frac{\dot{U}_{_{\pm }}\left( t\right) }{U_{_{\pm
}}\left( t\right) }=\frac{\dot{q}_{_{\pm }}\left( t\right) }{q_{_{\pm
}}\left( t\right) }+\alpha q_{_{\pm }}\left( t\right) \pm \zeta \text{ },
\label{NL1}
\end{equation}%
where $\alpha ,\zeta \in 
\mathbb{R}
  $ are positive constants to be determined in the linearization process. The constant term $\pm \zeta $ would allow us to indulge and study the effect of both dissipative and anti-dissipative  \cite{Bender 2016} (boosters/anti-damping) external forces as viable and realistic characterizations for dynamical systems. In the case of dissipative forces the phase-space trajectories shrink, whereas in the case of anti-dissipative forces the phase-space trajectories expand, as the dynamical systems evolve in time (e.g., \cite{Chandrasekar 2007,Mustafa PS 2021}). One should, hereby, be aware that we take $U_{_{+}}(t), q_{_{+}}(t)$ for $ +\zeta$ and $U_{_{-}}(t), q_{_{-}}(t)$ for $-\zeta$. 
  
Throughout we consider the motion of a classical particles with mass $m_{\circ }=1$ moving under the influence of a conservative quartic anharmonic potential force field%
\begin{equation}
V\left( q\right) =\frac{1}{2}\omega ^{2}q^{2}+\frac{1}{4}\alpha ^{2}q^{4}
\label{Duffing potential}
\end{equation}%
(often called the Duffing oscillator \cite{Amore 2005}) and subjected to a non-conservative dissipative (i.e, $\mathcal{R}_{_{+}}(q,\dot{q})$) and/or anti-dissipative (i.e., $\mathcal{R}_{_{-}}(q,\dot{q})$) Rayleigh force fields%
\begin{equation}
\mathcal{R}_{\pm}\left( q,\dot{q}\right) =\frac{1}{2}\left[ 3\alpha q\pm \gamma
_{_{1}}\left( \zeta \right) \right] \dot{q}^{2}+\left[ \gamma _{_{0}}\left(
\zeta \right) q\pm \gamma _{_{1}}\left( \zeta \right) \alpha q^{2}\right]\dot{q}.
\label{Rayleigh-0}
\end{equation}%
Then the standard Lagrangian describing this particle is given by
\begin{equation}
L\left( q,\dot{q},t\right) =\frac{1}{2}\dot{q}^{2}-\left[ \frac{1}{2}\omega
^{2}q^{2}+\frac{1}{4}\alpha ^{2}q^{4}\right], \label{L0}
\end{equation}%
and the corresponding Euler-Lagrange dynamical equation reads%
\begin{equation}
\frac{d}{dt}\left( \frac{\partial L}{\partial \dot{q}}\right) -\frac{%
\partial L}{\partial q}+\frac{\partial \mathcal{R}}{\partial \dot{q}}%
=0,  \label{EL0}
\end{equation}%
to cast the related dynamical equations%
\begin{equation}
\ddot{q}+\left[ 3\alpha\,q\pm\gamma_{_{1}}(\zeta)\right] \,\dot{q}%
+[\omega^2+\gamma_{_0}(\zeta)]q\pm \alpha\gamma_{_{1}}(\zeta)q^{2}+\alpha^{2}q^{3}=0,  \label{GMEE01}
\end{equation}%
to be linearized.  In the process, we shall use reverse engineering to come out with exact solutions for the HO and for the DHO using the most general solutions for the nonlinear GMEEs of (\ref{GMEE01}). In so doing, we would be imposing no initial conditions on the general solutions of (\ref{GMEE01}) so that such solutions would be the most explicit and general solutions to be adopted. Consequently, one would then use the appropriate initial conditions that suit the problem at hand.

The organization of the current methodical proposal is in order. In section 2, we consider the linearization of some GMEE-type (\ref{GMEE01}) dynamical equations into HO (with $\gamma_{_{0}}(\zeta)=\zeta^2$ and $\gamma_{_{1}}(\zeta)=2\zeta$) and DHO type ones (with $\gamma_{_{0}}(\zeta)=3\zeta^2$ and $\gamma_{_{1}}(\zeta)=4\zeta$) and report their exact solutions. Where, the details of our GMEE solutions suggest that they are explicit,  general and valid solutions not only for the GMEE-type equations (\ref{GMEE01}) but also for the HO and DHO type ones. No surprises for the general solution for the HO are observed. However, for the DHO we obtain the most explicit and general solution that offers consistency and comprehensive coverage for the associated under-damping, critical-damping, and over-damping cases. That is, no complex settings for $q_{_{\pm}}(t)$ and/or $\dot{q}_{_{\pm}}(t)$ are eminent/feasible. In section 3, moreover, we consider the linearization of an MEE-type (i.e., see (\ref{GMEE2}) below, which is a special case of (\ref{GMEE01}) with $\gamma_{_{0}}(\zeta)=-\zeta^2$ and $\gamma_{_{1}}(\zeta)=0$) dynamical equations into DHO-type ones and report their exact solutions. Therein, the reported solution turns out to be explicit, general and valid for the MEE-type (i.e., $q_{_{-}}(t)$) and for the DHO-type (i.e., $U_{_{-}}(t)$) equations. Moreover, for each dynamical system, we report illustrative figures for each $q_{_{\pm}}(t)$ and $\{ q_{_{\pm}}(t), p_{_{\pm}}(t)\}$ classical state (i.e., phase-space trajectory) as they evolve in time. To the best of our knowledge, the reported solutions, in both section 2 and 3, have never been reported elsewhere. Section 4 is devoted for our concluding remarks. 

\section{Linearization of GMEE-type into HO and DHO}

\subsection{Linearization into HO}

Consider the linear harmonic oscillator dynamical equation%
\begin{equation}
\ddot{U}\left( t\right) +\omega ^{2}U\left( t\right) =0.  \label{HO-eq}
\end{equation}%
A nonlocal transformation in the form of (\ref{NL1}) would transform (\ref{HO-eq}) into%
\begin{equation}
\ddot{q}_{_{\pm }}\left( t\right) +\left( 3\alpha q_{_{\pm }}\left( t\right)
\pm 2\zeta \right) \;\dot{q}_{_{\pm }}\left( t\right) +\text{ }\Omega
^{2}\,q_{_{\pm }}\left( t\right) \pm 2\zeta \alpha q_{_{\pm }}\left(
t\right) ^{2}+\alpha ^{2}q_{_{\pm }}\left( t\right) ^{3}=0\text{ };\text{ }%
\Omega ^{2}=\zeta ^{2}+\omega ^{2}.  \label{GMEE1}
\end{equation}%
This dynamical equation is the GMEE (\ref{GMEE01}) with $\gamma_{_{0}}(\zeta)=-\zeta^{2}$ and $\gamma_{_{1}}(\zeta)=0$). It describes a classical particle, with mass $m_{\circ}=1$, moving under the influence of a quartic anharmonic oscillator potential force field%
\begin{equation}
V\left( q_{_{\pm }}\right) =\frac{1}{2}\omega ^{2}q_{_{\pm }}^{2}+\frac{1}{4}%
\alpha ^{2}q_{_{\pm }}^{4},  \label{quartic HO-potential}
\end{equation}%
and subjected to the Rayleigh dissipative and/or anti-dissipative  force field%
\begin{equation}
\mathcal{R}_{_{\pm }}\left( q_{_{\pm }},\dot{q}_{_{\pm }}\right) =\frac{1}{2}\left(
3\alpha q_{_{\pm }}\pm 2\zeta \right) \;\dot{q}_{_{\pm }}^{2}+\left( \zeta
^{2}\pm 2\zeta \alpha q_{_{\pm }}^{2}\right) \;\dot{q}_{_{\pm }}.  \label{R1}
\end{equation}%
Then the standard Lagrangian for such a system is%
\begin{equation}
L\left( q_{_{\pm }},\dot{q}_{_{\pm }};t\right) =\frac{1}{2}\;\dot{q}_{_{\pm
}}^{2}-\frac{1}{2}\omega ^{2}q_{_{\pm }}^{2}-\frac{1}{4}\alpha ^{2}q_{_{\pm
}}^{4},  \label{GMEE-L1}
\end{equation}%
\begin{figure}[!ht]  
\centering
\includegraphics[width=0.3\textwidth]{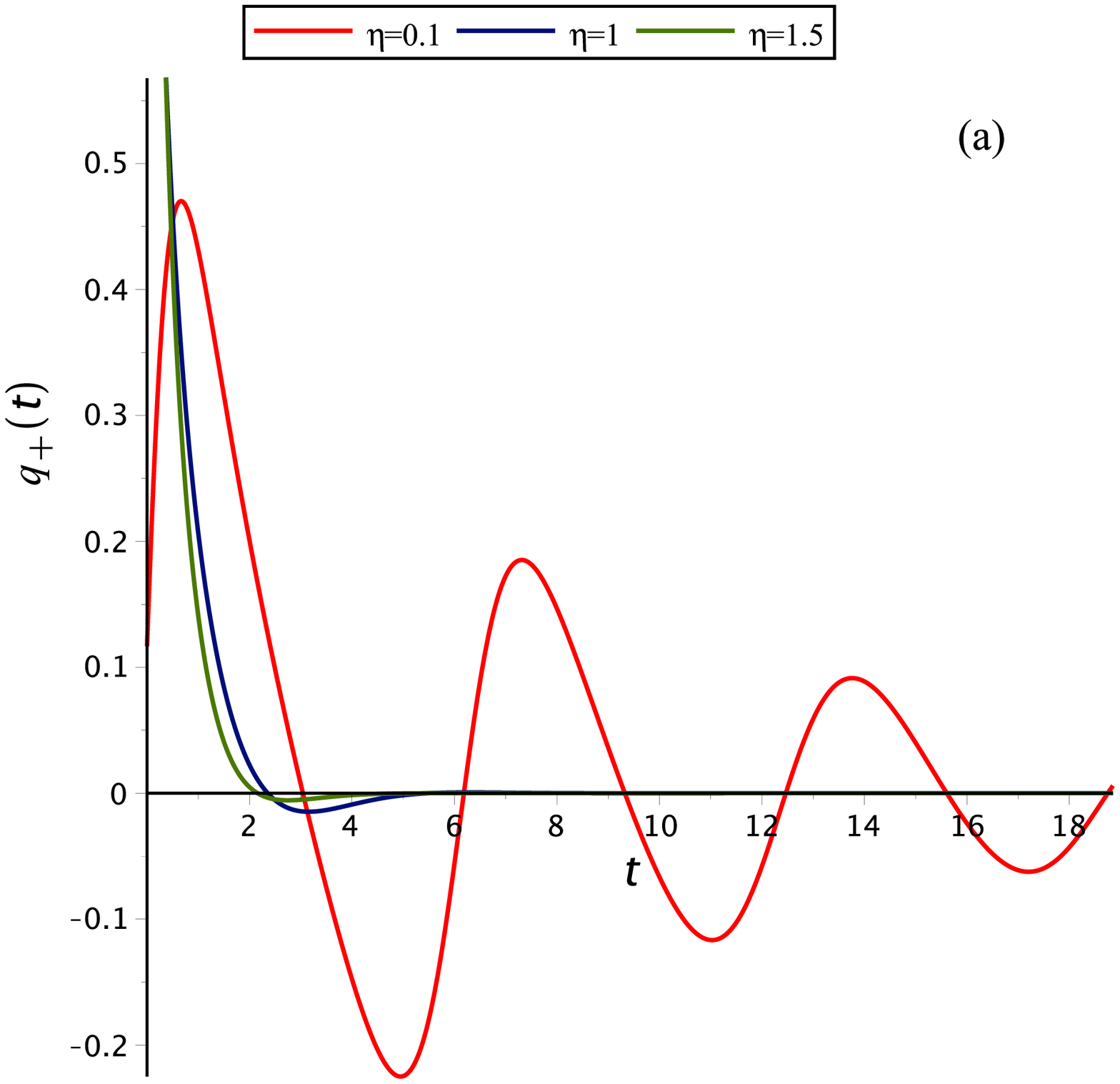}
\includegraphics[width=0.3\textwidth]{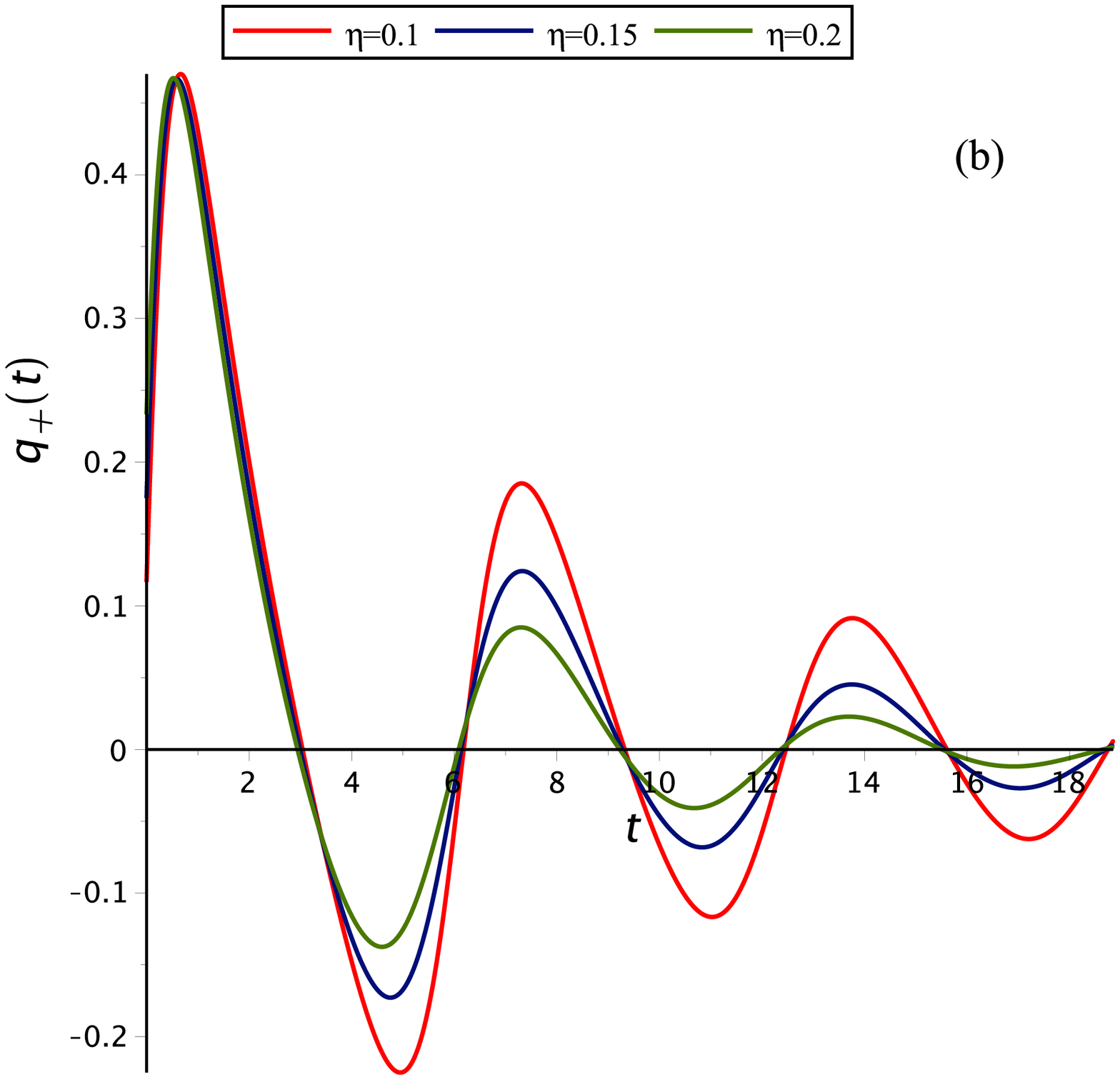}
\includegraphics[width=0.3\textwidth]{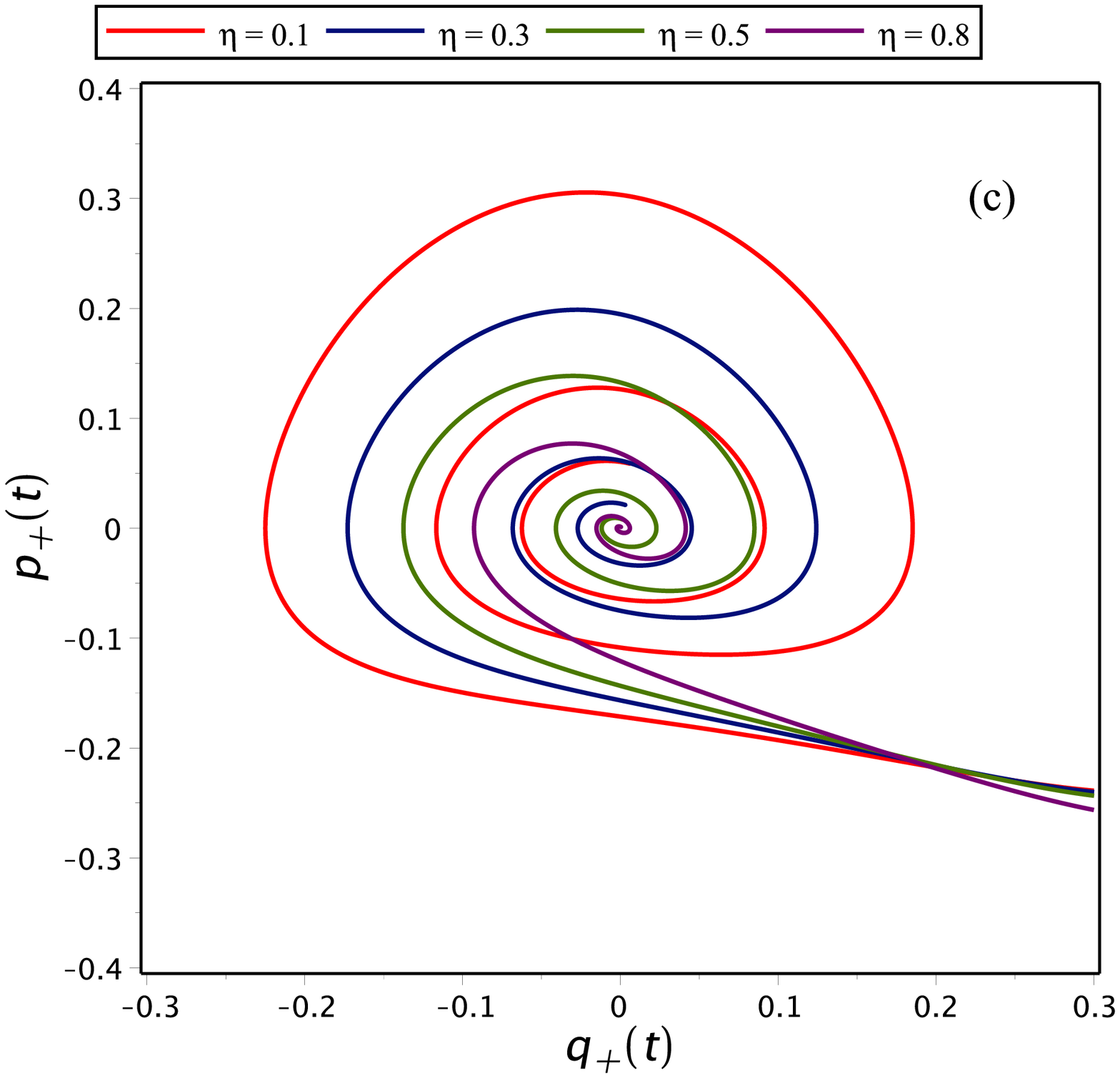}
\includegraphics[width=0.3\textwidth]{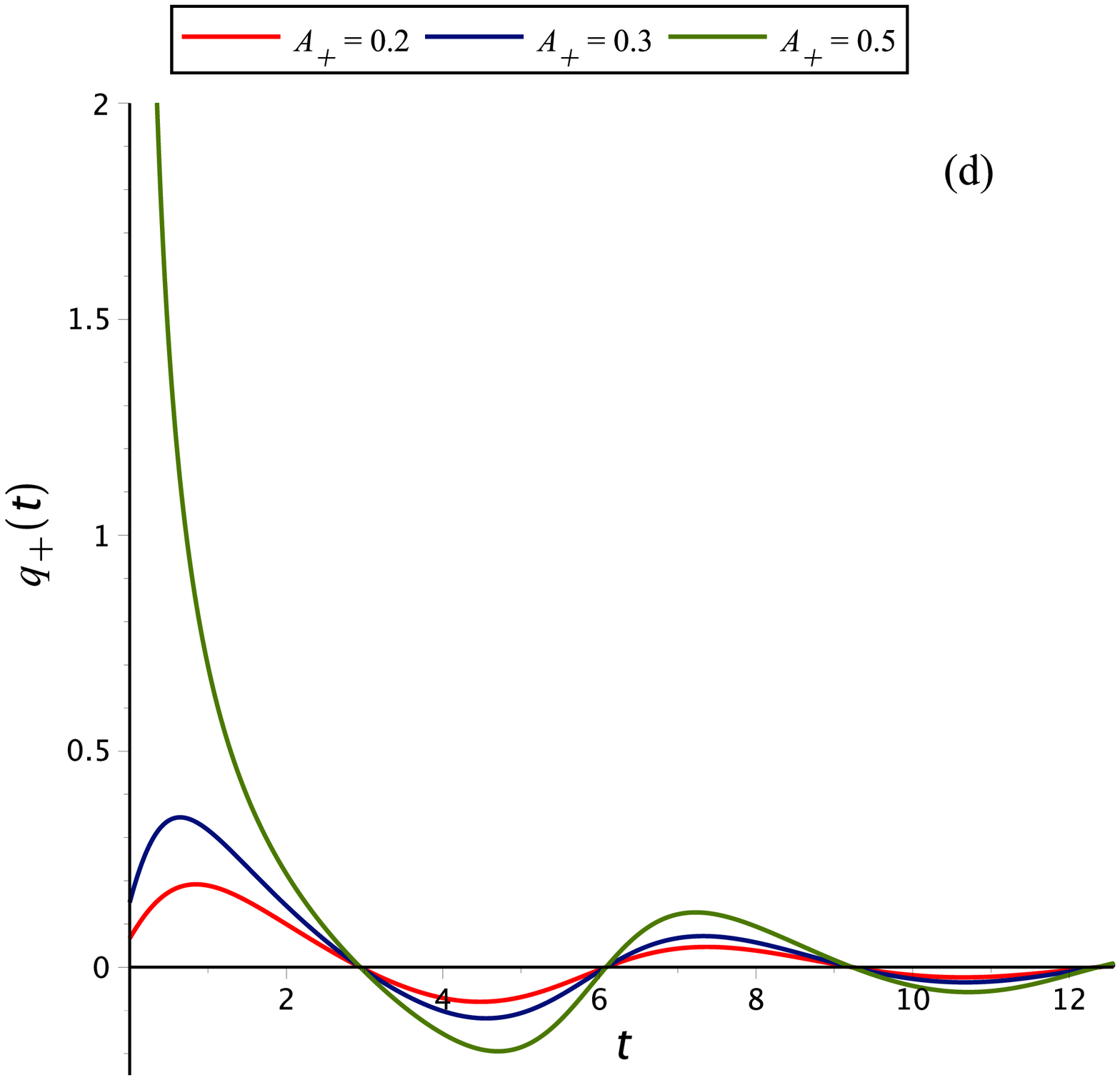}
\includegraphics[width=0.3\textwidth]{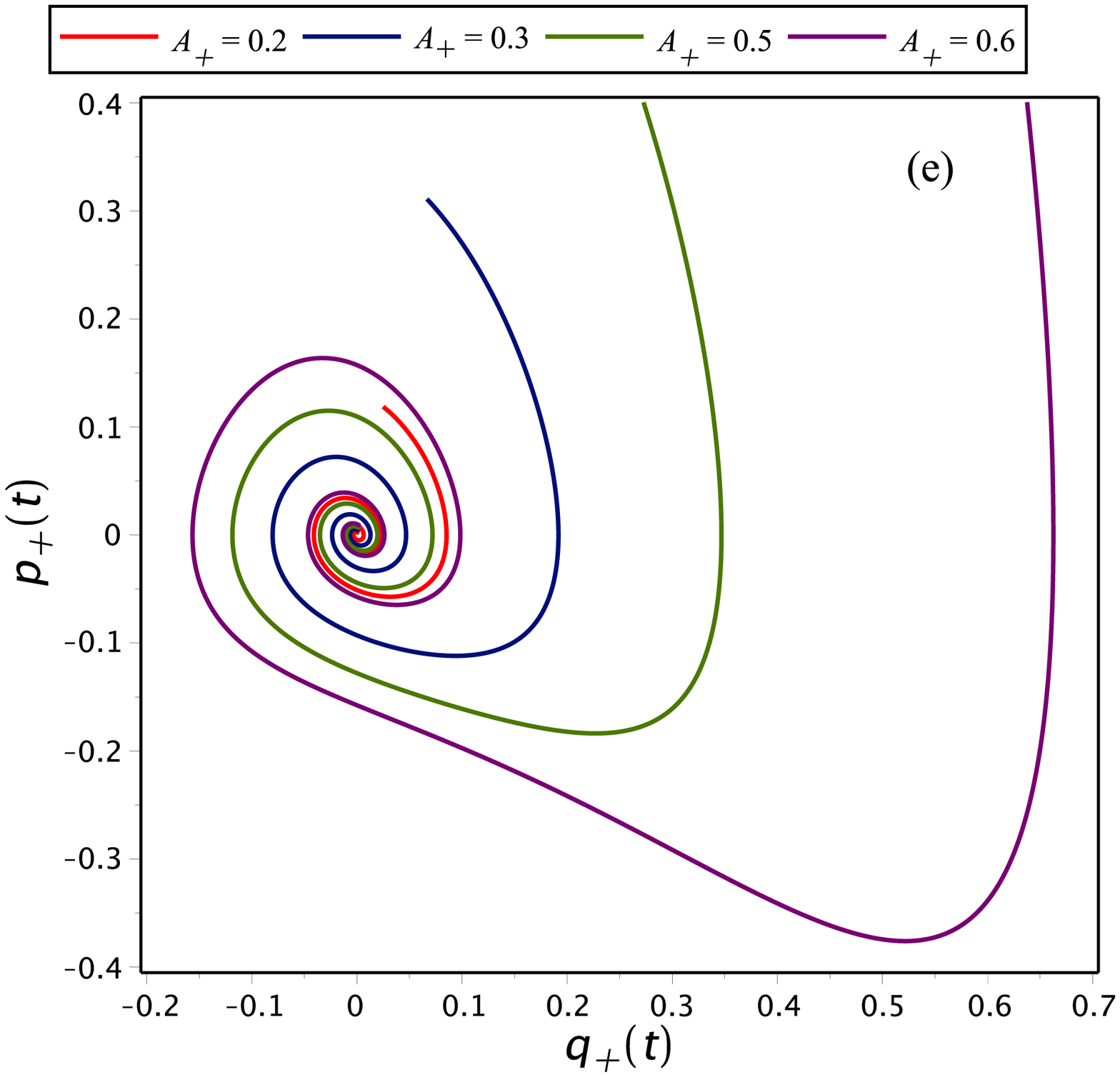} 
\includegraphics[width=0.3\textwidth]{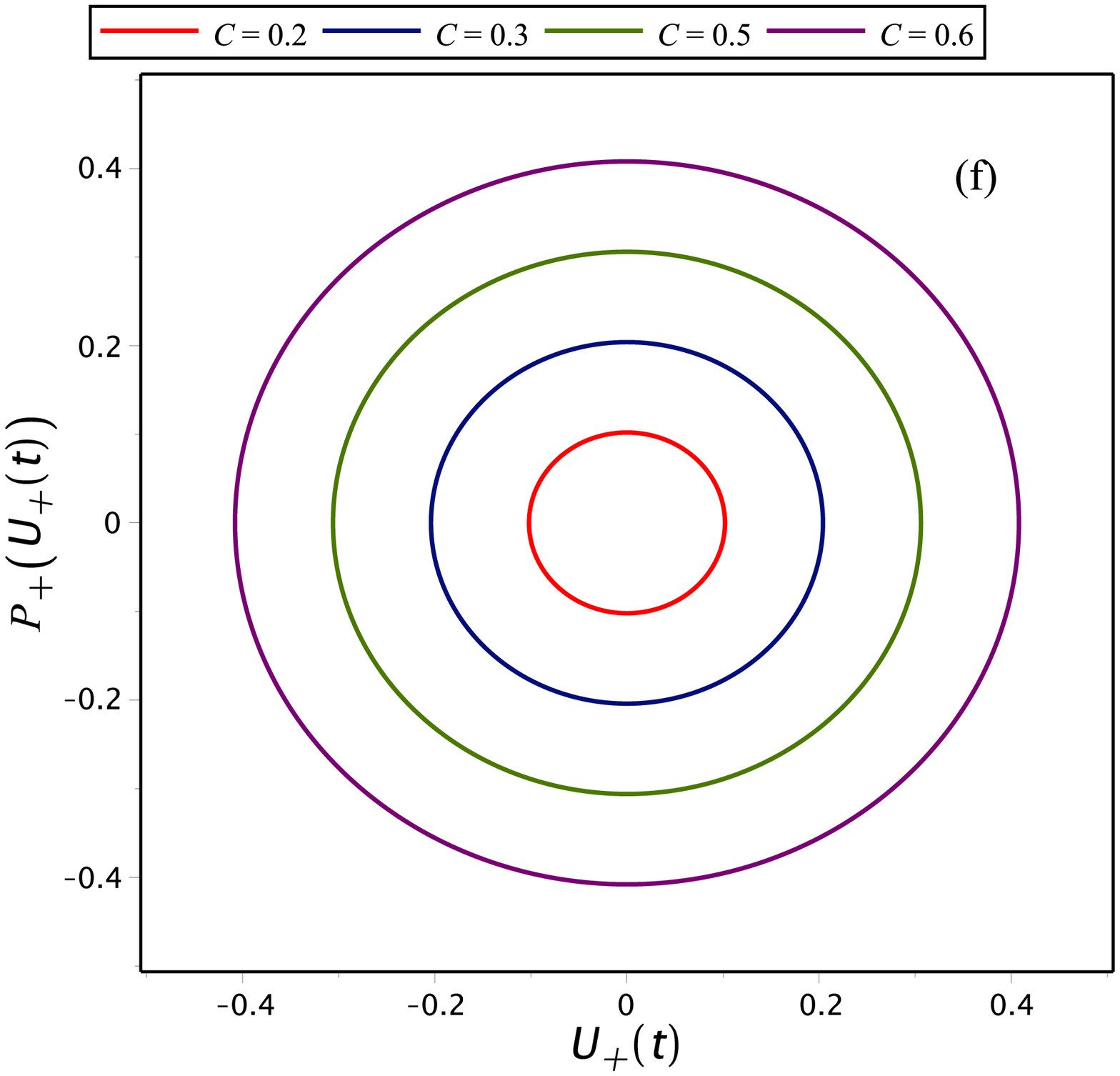}
\caption{\small 
{ For $A_{_{+}}=0.35$ and $\omega=1$ we show (a) $q_{_{+}}(t)$ of (\ref{GMEE1-sol}) for $\eta=0.1$ (under-damping), $\eta=1$ (critical-damping), and for $\eta=1.5$ (over-damping), (b) $q_{_{+}}(t)$ of (\ref{GMEE1-sol}) for different values of $\eta<1$, and (c) $\{q_{_{+}}(t),p_{_{+}}(t)=\dot{q}_{_{+}}(t)\}$ of (\ref{GMEE1-sol}) for different values of $\eta<1$. For $\eta=0.2$ and $\omega=1$ we show (d) $q_{_{+}}(t)$ of (\ref{GMEE1-sol}) for different values of $A_{_{+}}$, (e) phase-space trajectories for $\{q_{_{+}}(t),p_{_{+}}(t)=\dot{q}_{_{+}}(t)\}$ of (\ref{GMEE1-sol}) for different values of $A_{_{+}}$, and (f) phase-space trajectories $\{U_{_{+}}(t),P_{_{+}}(U_{_{+}})=\dot{U}_{_{+}}(t)\}$ of (\ref{HO-sol}) for different values of $C$.}}
\label{fig1}
\end{figure}%
and the corresponding dynamical equation (\ref{GMEE1}) is obtained by using
the Euler-Lagrange recipe%
\begin{equation}
\frac{d}{dt}\left( \frac{\partial L}{\partial \dot{q}_{_{\pm }}}\right) -%
\frac{\partial L}{\partial q_{_{\pm }}}+\frac{\partial \mathcal{R}_{_{\pm }}%
}{\partial \dot{q}_{_{\pm }}}=0.  \label{Euler-Lagrange}
\end{equation}%
Equation (\ref{GMEE1}) admits an exact solution in the form of%
\begin{equation}
q_{_{\pm }}\left( t\right) =\mp \frac{1}{\alpha }\left\{ \frac{C_{2}\;\left(
\zeta \mp i\omega \right) \;e^{i\omega t}+\left( \zeta \pm i\omega \right)
\;e^{-i\omega t}}{C_{1}+\left( C_{2}\;e^{i\omega t}+\;e-^{i\omega t}\right)
e^{\mp \zeta t}}\right\} e^{\mp \zeta t}.  \label{GMEE1-sol-0}
\end{equation}%
Which, in a straightforward manner, with the assumptions (to facilitate the linearization process) that $C_{2}=1$, $\alpha =2\omega $, $\zeta =\omega
\eta $ and $A_{_{\pm }}=\mp 1/C_{1}$, could be simplified to read%
\begin{equation}
q_{_{\pm }}\left( t\right) =A_{_{\pm }}\left\{ \frac{\eta \cos \left( \omega
t\right) \pm \sin \left( \omega t\right) }{1\mp 2A_{_{\pm }}e^{\mp \omega
\eta t}\cos \left( \omega t\right) }\right\} e^{\mp \omega \eta t}.
\label{GMEE1-sol}
\end{equation}%
This would in turn imply, through the nonlocal transformation (\ref{NL1}),
that%
\begin{equation}
U_{_{\pm }}\left( t\right) =C\left[ \eta \cos \left( \omega t\right) \pm
\sin \left( \omega t\right) \right]   \label{HO-sol}
\end{equation}%
as the exact textbook solution for the harmonic oscillator in (\ref{HO-eq}), where one would set $a=C\eta $ and $b=\pm C$. However, $U_{_{\pm }}\left(
t\right) $ of (\ref{HO-sol}) is the solution for the harmonic oscillator (\ref{HO-eq}) that provides a consistent nonlocal connection, through the nonlocal transformation in (\ref{NL1}), to the generalized modified Emden
equation (\ref{GMEE1}) along with its exact general solution (\ref{GMEE1-sol}). The linearization of the GMEE of (\ref{GMEE1}) into HO (\ref{HO-eq}) is clear, therefore.%
\begin{figure}[!ht]  
\centering
\includegraphics[width=0.2\textwidth]{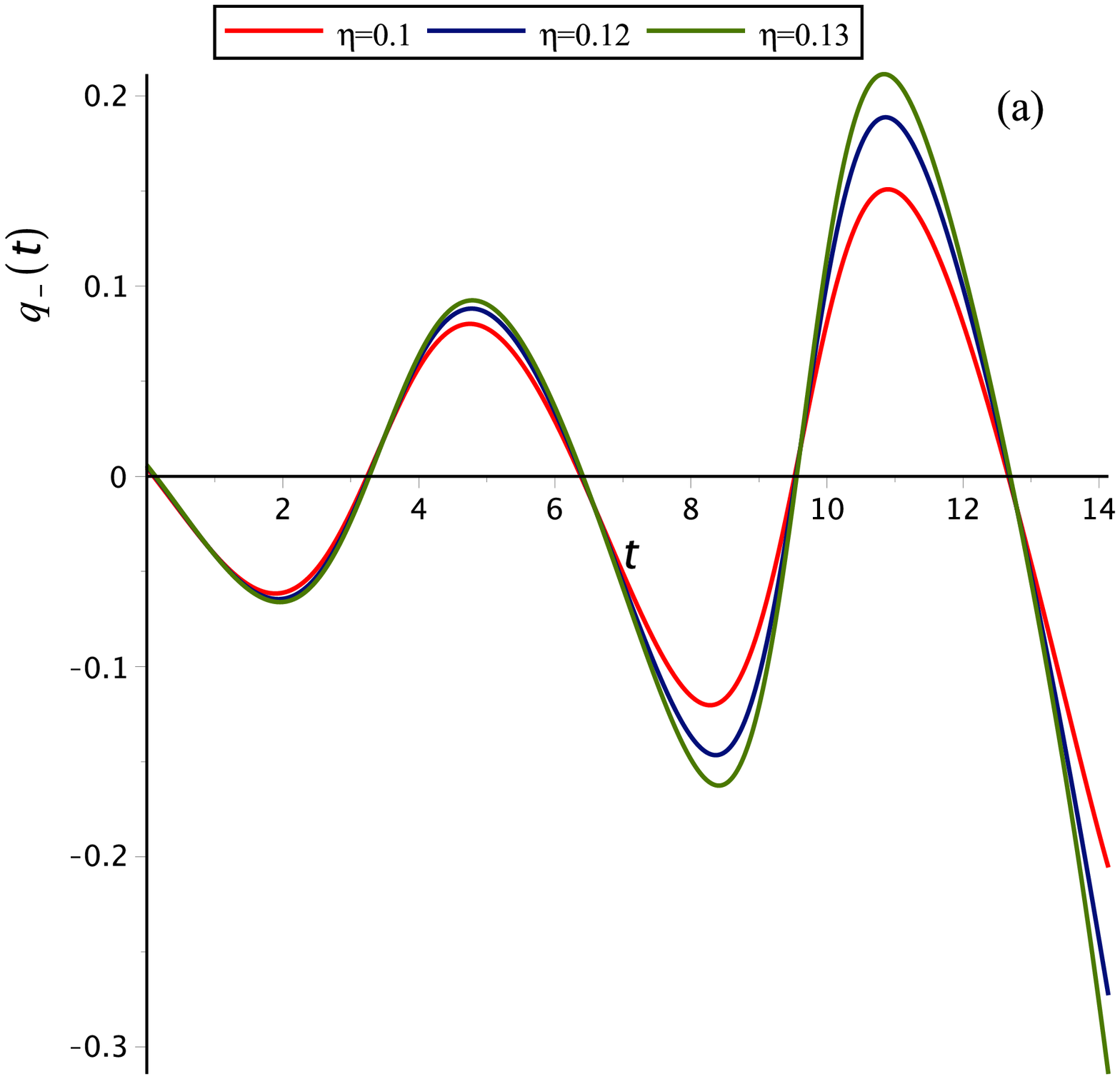}
\includegraphics[width=0.2\textwidth]{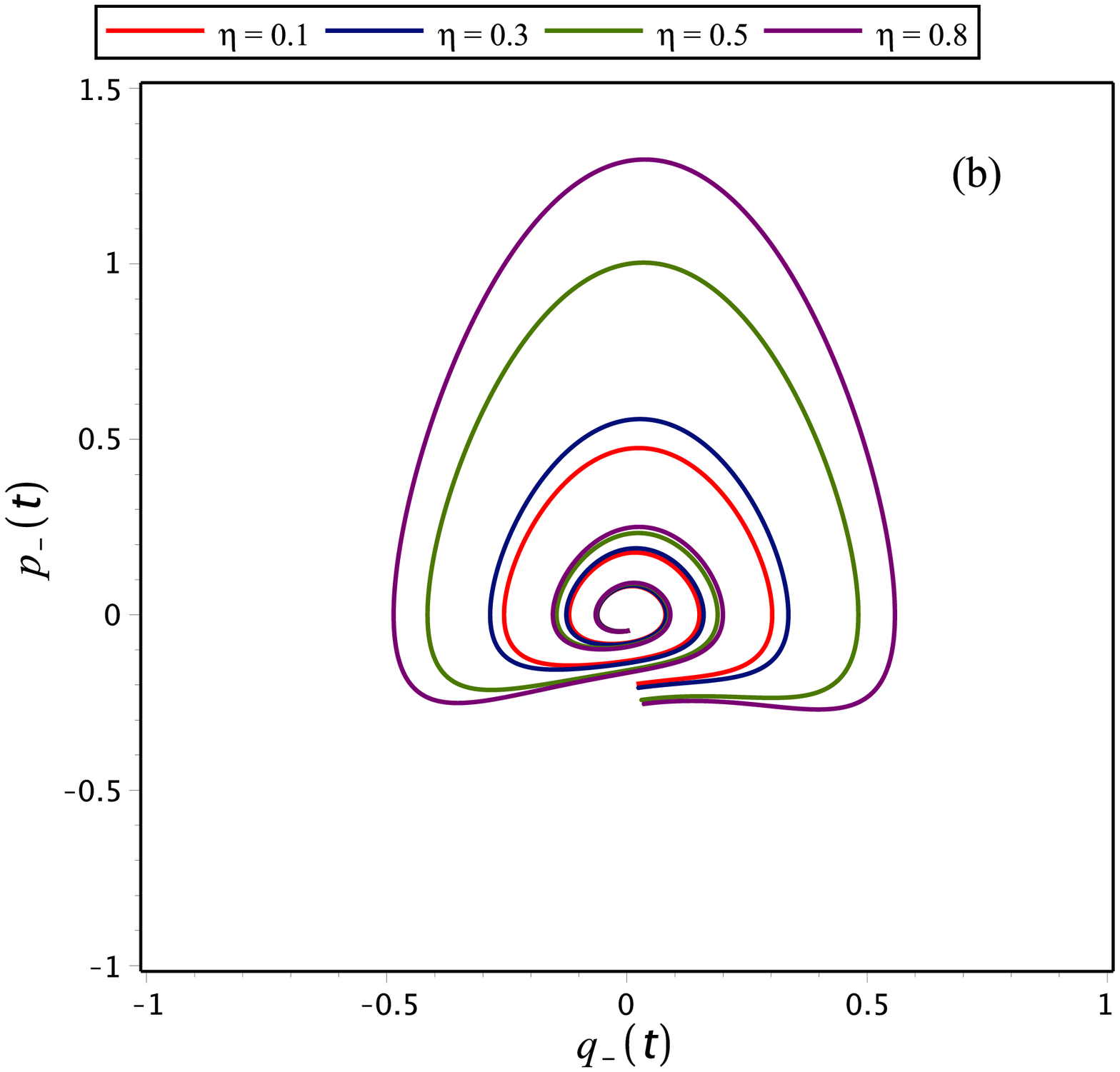} 
\includegraphics[width=0.2\textwidth]{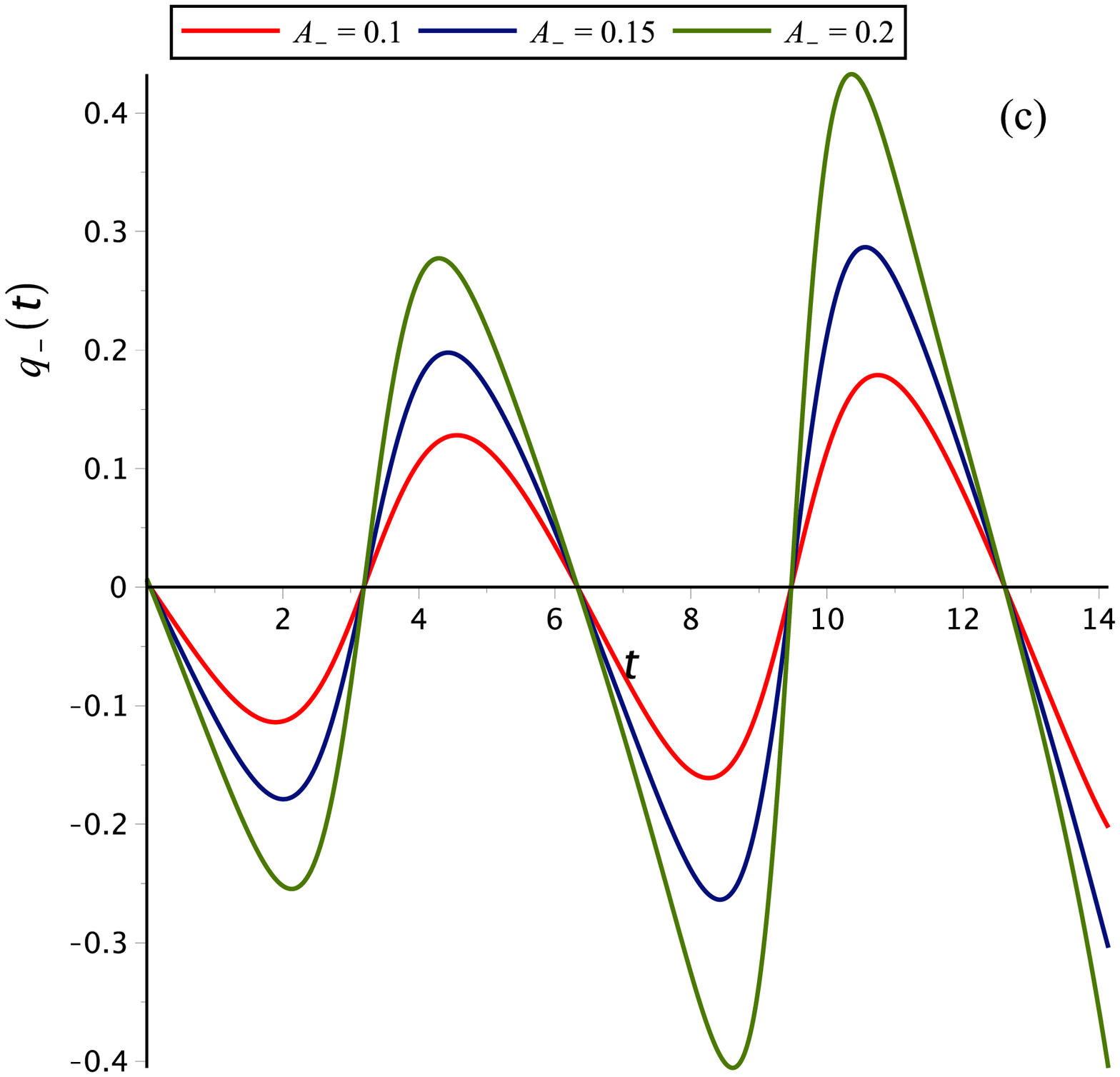}
\includegraphics[width=0.2\textwidth]{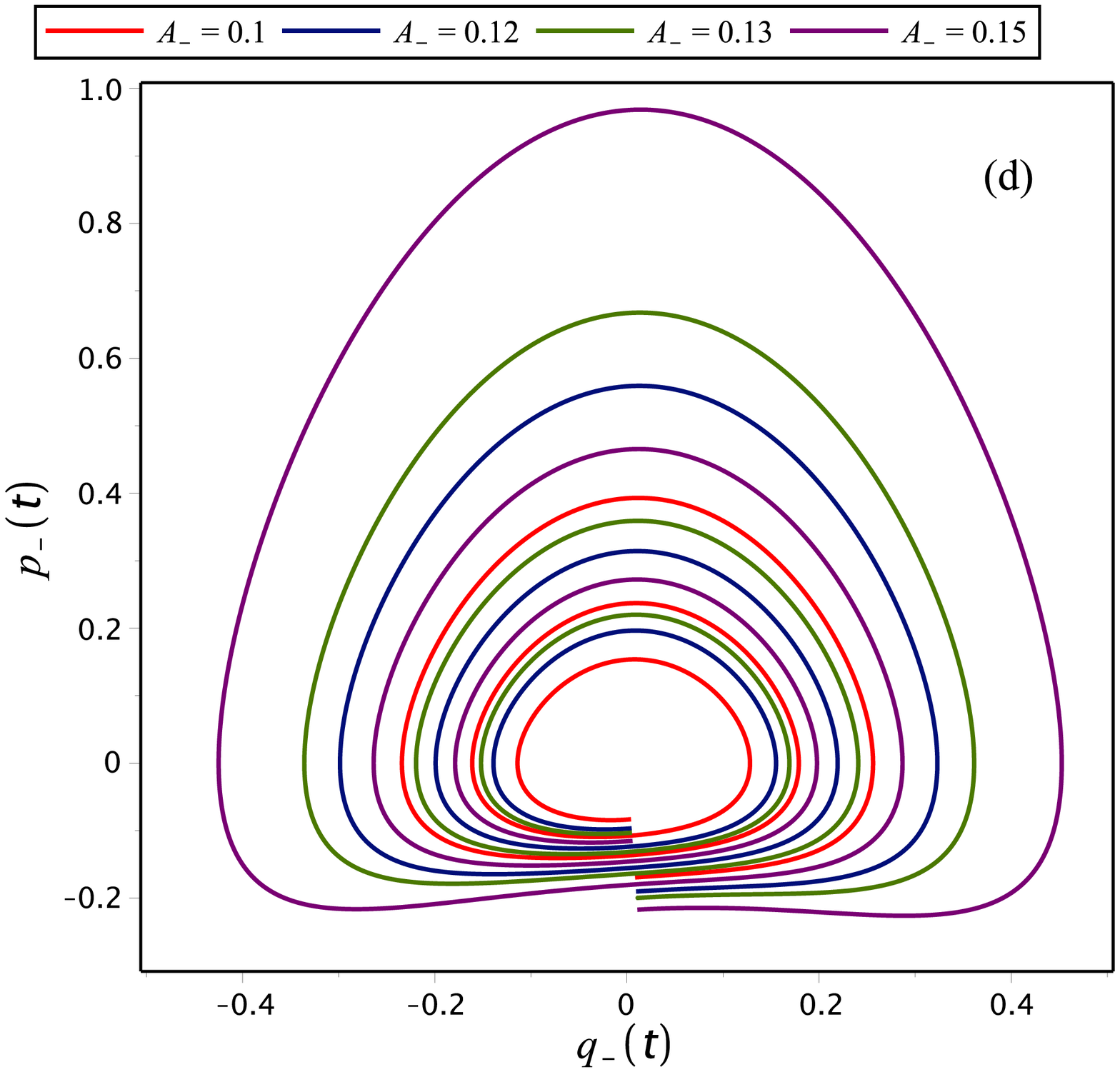}
\caption{\small 
{ For $A_{_{-}}=0.05$, and $\omega=1$ we show (a) $q_{_{-}}(t)$ of (\ref{GMEE1-sol}) for different values of $\eta<1$, (b) phase-space trajectories for $\{q_{_{-}}(t),p_{_{-}}(t)=\dot{q}_{_{-}}(t)\}$ of (\ref{GMEE1-sol}) for different $\eta<1$ values. For $\eta=0.05$, and $\omega=1$ we show (c) $q_{_{-}}(t)$ of (\ref{GMEE1-sol}) for different $A_{_{-}}$ values, and (d) phase-space trajectories $\{q_{_{-}}(t),p_{_{-}}(t)=\dot{q}_{_{-}}(t)\}$ of (\ref{GMEE1-sol}) for different values of $A_{_{-}}$.}}
\label{fig2}
\end{figure}%

In Figure 1, we use $\mathcal{R}_{_{+}}(q,\dot{q})$ of (\ref{R1}) (i.e., only dissipative effects of the Rayleigh force field are involved) and plot,  (a) $q_{_{+}}(t)$ of (\ref{GMEE1-sol}) for $\eta=0.1<1$ (under damping), $\eta=1$ (critical damping) and $\eta=1.5>1$ (over damping), as they evolve in time. One immediately observes that such a system exactly resembles the DHO behaviour. This DHO resemblance is inherited from the nonlocal connection $U_{_{+}}(t)$ of (\ref{NL1}) between the HO (\ref{HO-eq}) and the GMEE (\ref{GMEE1}). In (b) and (d), we show $q_{_{+}}(t)$ for different values of $\eta<1$ and $A_{_{+}}$, respectively, as they evolve in time. It is clear that the frequencies of oscillations are isochronic and amplitude-independent. The phase-space trajectories (i.e., the classical states trajectories $\{q_{_{+}}(t),p_{_{+}}(t)\}$) are plotted in (c) and (e) for different values of $\eta<1$ and $A_{_{+}}$, respectively, as they evolve in time. The trajectories exemplify the DHO behaviour as they clockwise shrink/decrease. In (f) we show the phase-space trajectory $\{U_{_{+}}(t),P_{_{+}}(U_{_{+}})=\dot{U}_{_{+}}(t)\}$ of (\ref{HO-sol}) for different amplitude $C$ values. They are indeed the textbook phase-space trajectories for the HO (\ref{HO-eq}).

In Figure 2, moreover, we show in (a) and (c) the effect of the anti-dissipative force fields $\mathcal{R}_{_{-}}(q_{_{-}},\dot{q}_{_{-}})$ (\ref{R1}) on $q_{_{-}}(t)$ as it evolves in time for different values of $\eta<1$ and $A_{_{-}}$, respectively. We clearly observe that the effective amplitude of oscillation increases with time while performing isochronous oscillations. In (b) and (c) we show the phase-space trajectories as they evolve in time for different values of $\eta<1$ and $A_{_{-}}$, respectively. One may observe that such trajectories are clockwise expanding as they evolve in time. A common characteristic behaviour for dynamical systems exposed to anti-dissipative Rayleigh force fields $\mathcal{R}_{_{-}}(q_{_{-}},\dot{q}_{_{-}})$ of (\ref{R1}).

\subsection{Linearization into DHO}

We now consider the damped harmonic oscillator equation%
\begin{equation}
\ddot{U}\left( t\right) +2\zeta \dot{U}\left( t\right) +\omega ^{2}U\left(
t\right) =0  \label{DHO-eq}
\end{equation}%
and use our nonlocal transformation recipe (\ref{NL1}) for $U_{_{+}}\left(
t\right) $ so that%
\begin{equation}
U_{_{+}}\left( t\right) =q_{_{+}}\left( t\right) \exp \left( \int \left(
\alpha q_{_{+}}\left( t\right) +\zeta \right) dt\right)   \label{U-DHO}
\end{equation}%
to obtain%
\begin{equation}
\ddot{q}_{_{+}}\left( t\right) +\left( 3\alpha q_{_{+}}\left( t\right)
+4\zeta \right) \;\dot{q}_{_{+}}\left( t\right) +\text{ }\tilde{\Omega}%
^{2}\,q_{_{+}}\left( t\right) +4\zeta \alpha q_{_{+}}\left( t\right)
^{2}+\alpha ^{2}q_{_{+}}\left( t\right) ^{3}=0\text{ };\text{ }\tilde{\Omega}%
^{2}=3\zeta ^{2}+\omega ^{2}.  \label{GMEE2}
\end{equation}%
\begin{figure}[!ht]  
\centering
\includegraphics[width=0.2\textwidth]{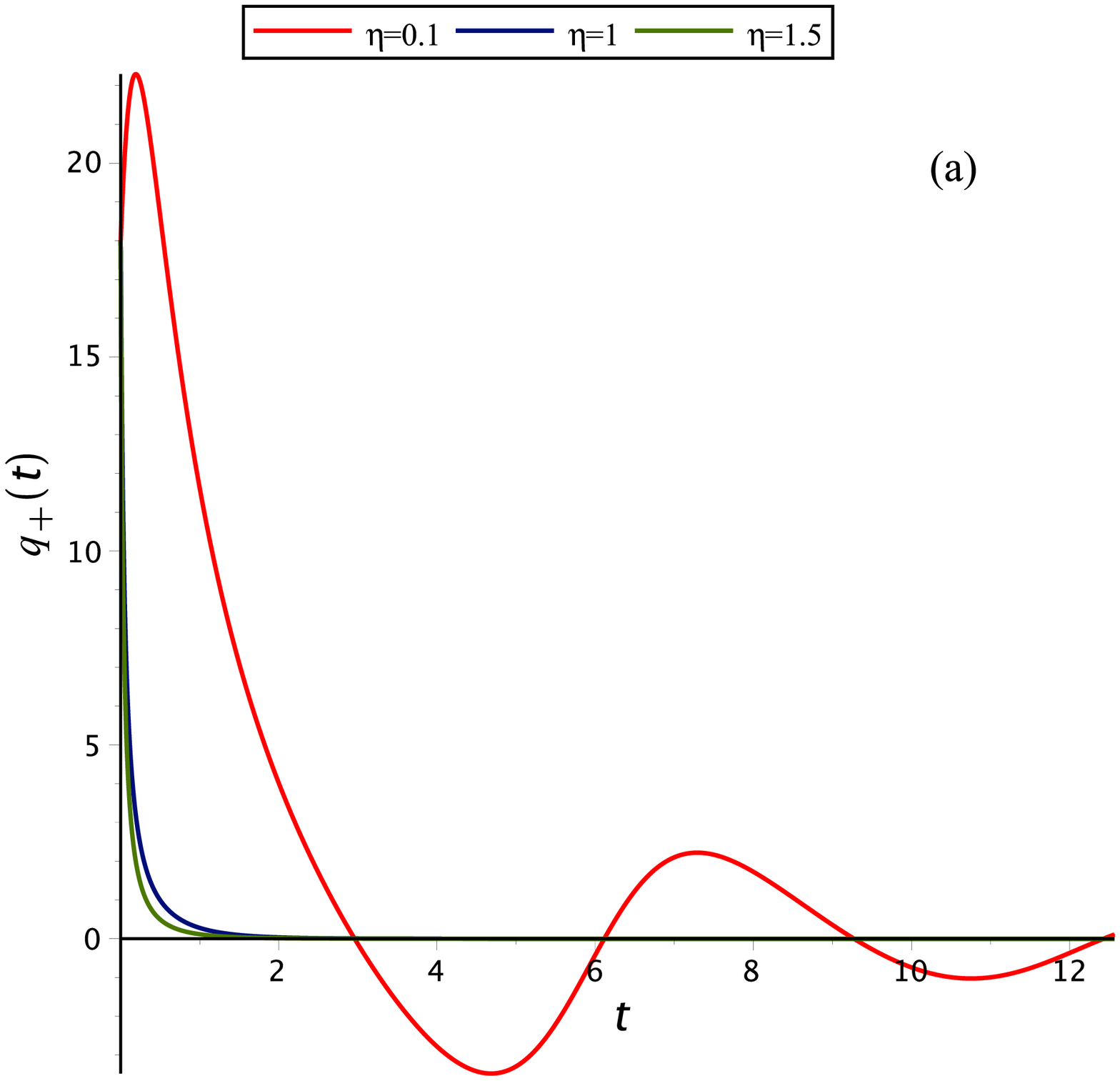}
\includegraphics[width=0.2\textwidth]{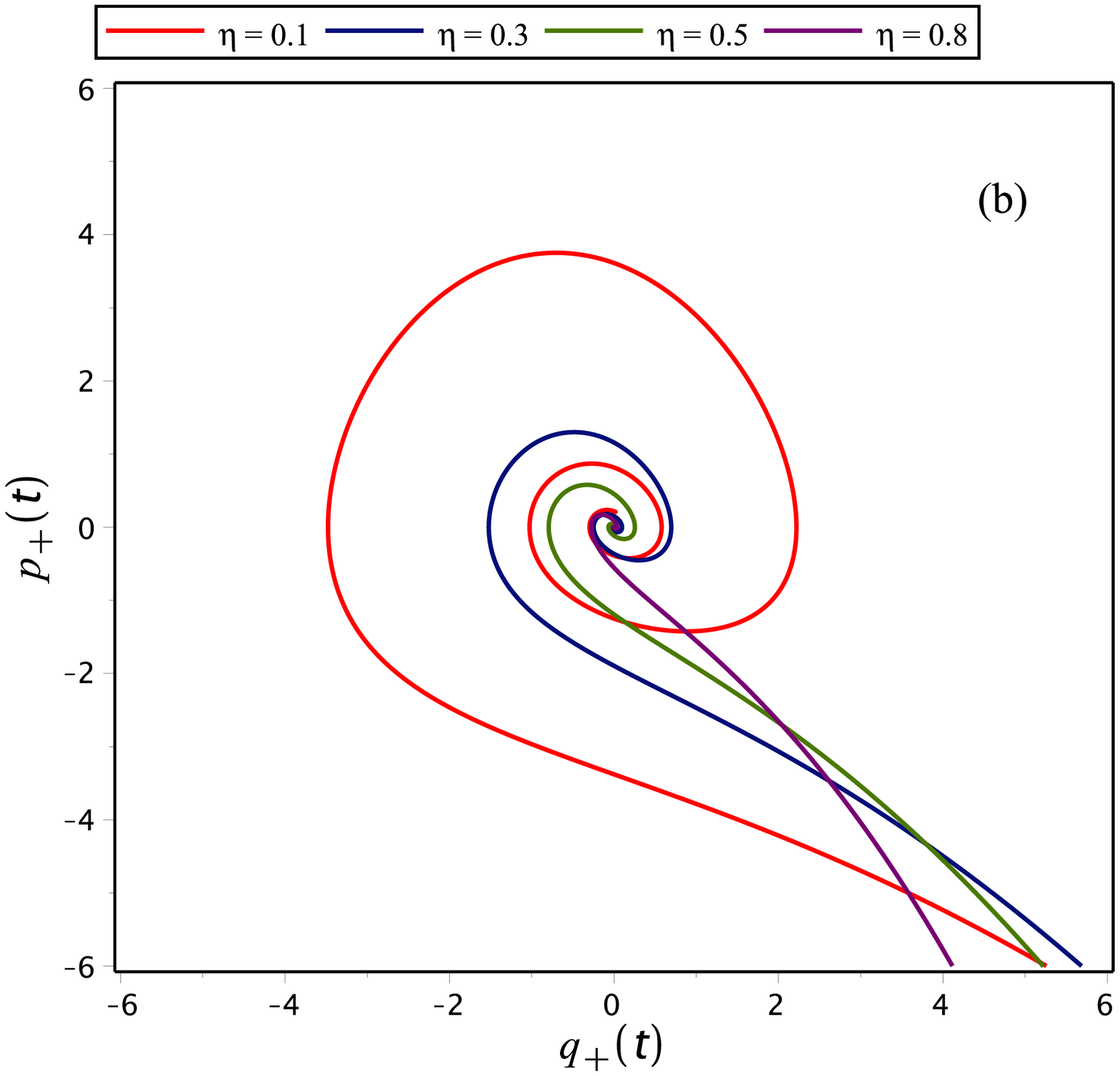} 
\includegraphics[width=0.2\textwidth]{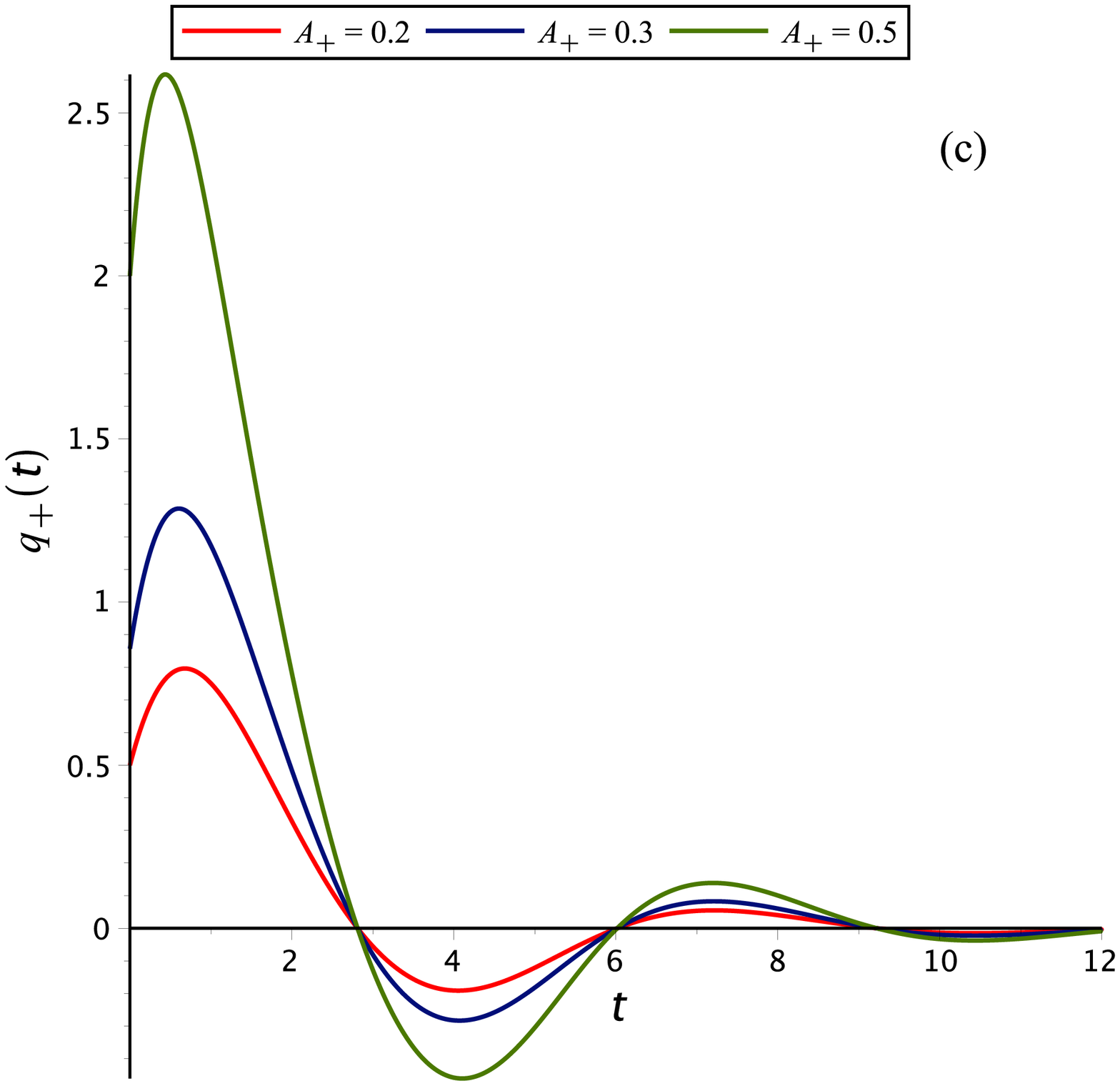}
\includegraphics[width=0.2\textwidth]{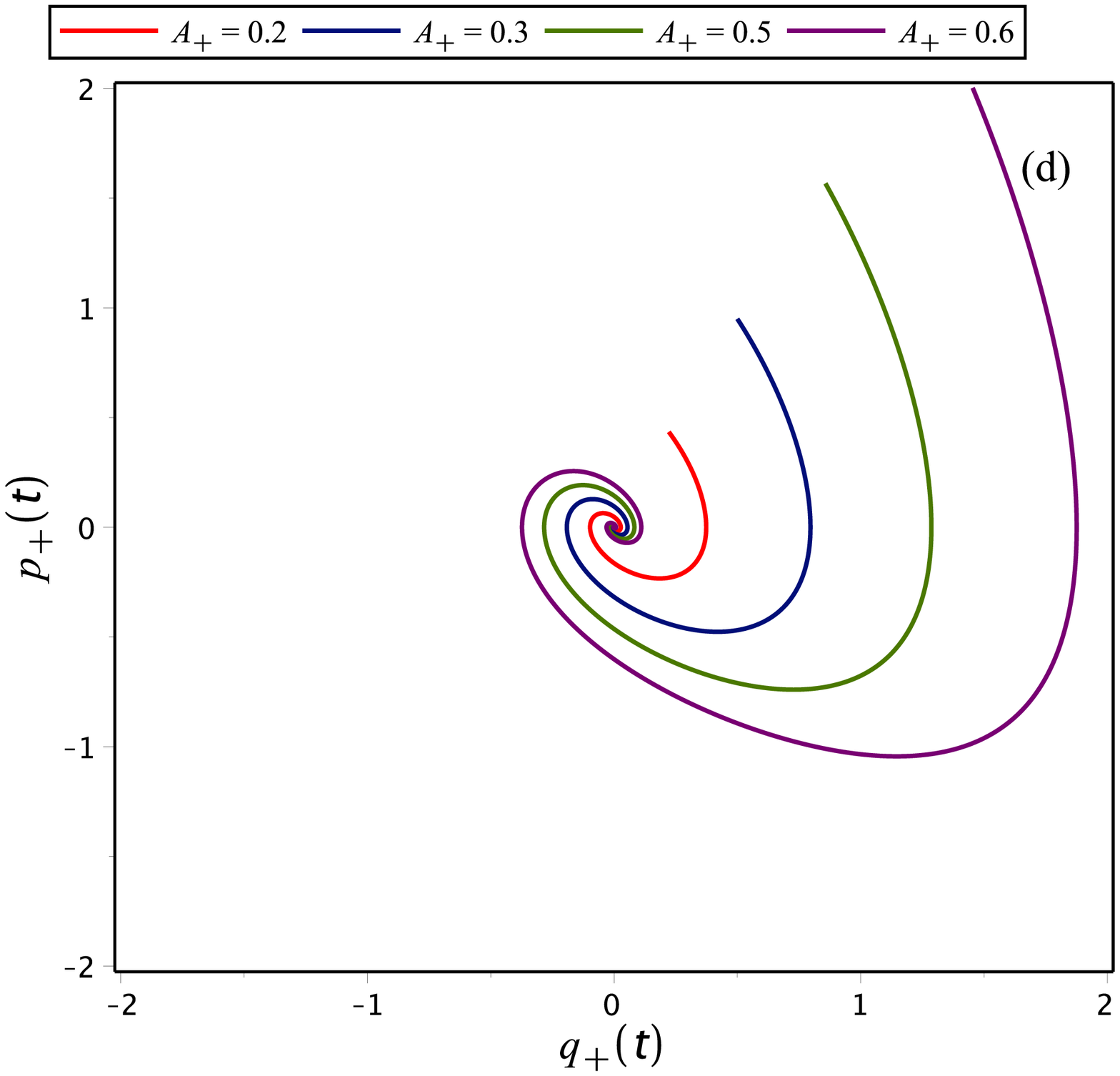}
\caption{\small 
{ For $A_{_{+}}=0.9$ and $\omega=1$ we show (a) $q_{_{+}}(t)$ of (\ref{GMEE2-sol}) for $\eta=0.1$ (under-damping), $\eta=1$ (critical-damping), and for $\eta=1.5$ (over-damping), (b) $\{q_{_{+}}(t),p_{_{+}}(t)=\dot{q}_{_{+}}(t)\}$ of (\ref{GMEE2-sol}) for different values of $\eta<1$. For $\eta=0.2$, $\omega=1$ and different values of $A_{_{+}}$, we show (c) $q_{_{+}}(t)$ of (\ref{GMEE2-sol}), and (d) phase-space trajectory for $\{q_{_{+}}(t),p_{_{+}}(t)=\dot{q}_{_{+}}(t)\}$ of (\ref{GMEE2-sol}).}}
\label{fig3}
\end{figure}%
\\
Such nonlinear dynamical equation (i.e.,  GMEE (\ref{GMEE01}) with $\gamma_{_{0}}(\zeta)=3\zeta^{2}$ and $\gamma_{_{1}}(\zeta)=4\zeta$) describes a classical particle, with mass $m_{\circ }=1$, moving under the influence of the quartic anharmonic oscillator potential force field of (\ref{quartic HO-potential}) and subjected to the Rayleigh dissipative force field%
\begin{equation}
\mathcal{R}_{_{+}}\left( q_{_{+}},\dot{q}_{_{+}}\right) =\frac{1}{2}\left( 3\alpha
q_{_{+}}\left( t\right) +4\zeta \right) \;\dot{q}_{_{+}}\left( t\right)
^{2}+\left( 3\zeta ^{2}+4\zeta \alpha q_{_{+}}\left( t\right) ^{2}\right) \;%
\dot{q}_{_{+}}\left( t\right) .  \label{R2}
\end{equation}%
Then the standard Lagrangian for such a system is readily given in (\ref{GMEE-L1}) and the corresponding nonlinear dynamical equation (\ref{GMEE2}) is obtained through the Euler-Lagrange recipe of (\ref{Euler-Lagrange}).
Nevertheless, the nonlinear dynamical equation (\ref{GMEE2}) admits an exact solution in the form of%
\begin{equation}
q_{_{+}}\left( t\right) =-\frac{e^{-2\zeta t}}{\alpha }\left\{ \frac{\left(
2\zeta +\beta \right) e^{-\beta t}+C_{2}\left( 2\zeta -\beta \right)
e^{\beta t}}{C_{1}+e^{-2\zeta t}\left( C_{2}e^{\beta t}+e^{-\beta t}\right) }%
\right\} \text{ };\text{ }\beta =\omega \sqrt{\eta ^{2}-1}.
\label{GMEE2-sol-0}
\end{equation}%
Which is simplified, with $C_{2}=1$, $\alpha =$ $\zeta =\omega \eta $, and $%
C_{1}=-2/A$, into%
\begin{equation}
q_{_{+}}\left( t\right) =A\left\{ \frac{2\cosh \left( \beta t\right) -\left(
\beta /\omega \eta \right) \sinh \left( \beta t\right) }{1-Ae^{-2\omega \eta
t}\cosh \left( \beta t\right) }\right\} e^{-2\omega \eta t}.
\label{GMEE2-sol}
\end{equation}%
This result when substituted in (\ref{U-DHO}) yields%
\begin{equation}
U_{+}\left( t\right) =A\left[ 2\cosh \left( \beta t\right) -\frac{\beta }{%
\omega \eta }\sinh \left( \beta t\right) \right] e^{-\omega \eta t},
\label{DHO-sol}
\end{equation}%
as the exact solution of (\ref{DHO-eq}). Then the linearization of (\ref {GMEE2}) into the linear damped harmonic oscillator equation (\ref{DHO-eq}) is clear.

In Figure 3, we show the linearization into DHO (\ref{DHO-eq}) effect on the GMEE of (\ref{GMEE2}) and plot (a) $q_{_{+}}(t)$ of (\ref{GMEE2-sol}) for $\eta=0.1<1$ (under damping), $\eta=1$ (critical damping) and $\eta=1.5>1$ (over damping), as they evolve in time, (b) $\{q_{_{+}}(t),p_{_{+}}(t)=\dot{q}_{_{+}}(t)\}$ of (\ref{GMEE2-sol}) for different values of $\eta<1$. For $\eta=0.2$, $\omega=1$ and different values of $A_{_{+}}$ we show (c) $q_{_{+}}(t)$ of (\ref{GMEE2-sol}), and (d) phase-space trajectory for $\{q_{_{+}}(t),p_{_{+}}(t)=\dot{q}_{_{+}}(t)\}$ of (\ref{GMEE2-sol}). Again, one observes that the system resembles the DHO behaviour, where the phase-space trajectories in (b) and (d) show similar trend of clockwise shrinking/decreasing as they evolve in time.

\section{Linearization of MEE-type into DHO}

We now consider the damped harmonic oscillator equation in (\ref{DHO-eq})
and use our nonlocal transformation recipe (\ref{NL1}) for $U_{-}\left(
t\right) $ so that%
\begin{figure}[!ht]  
\centering
\includegraphics[width=0.3\textwidth]{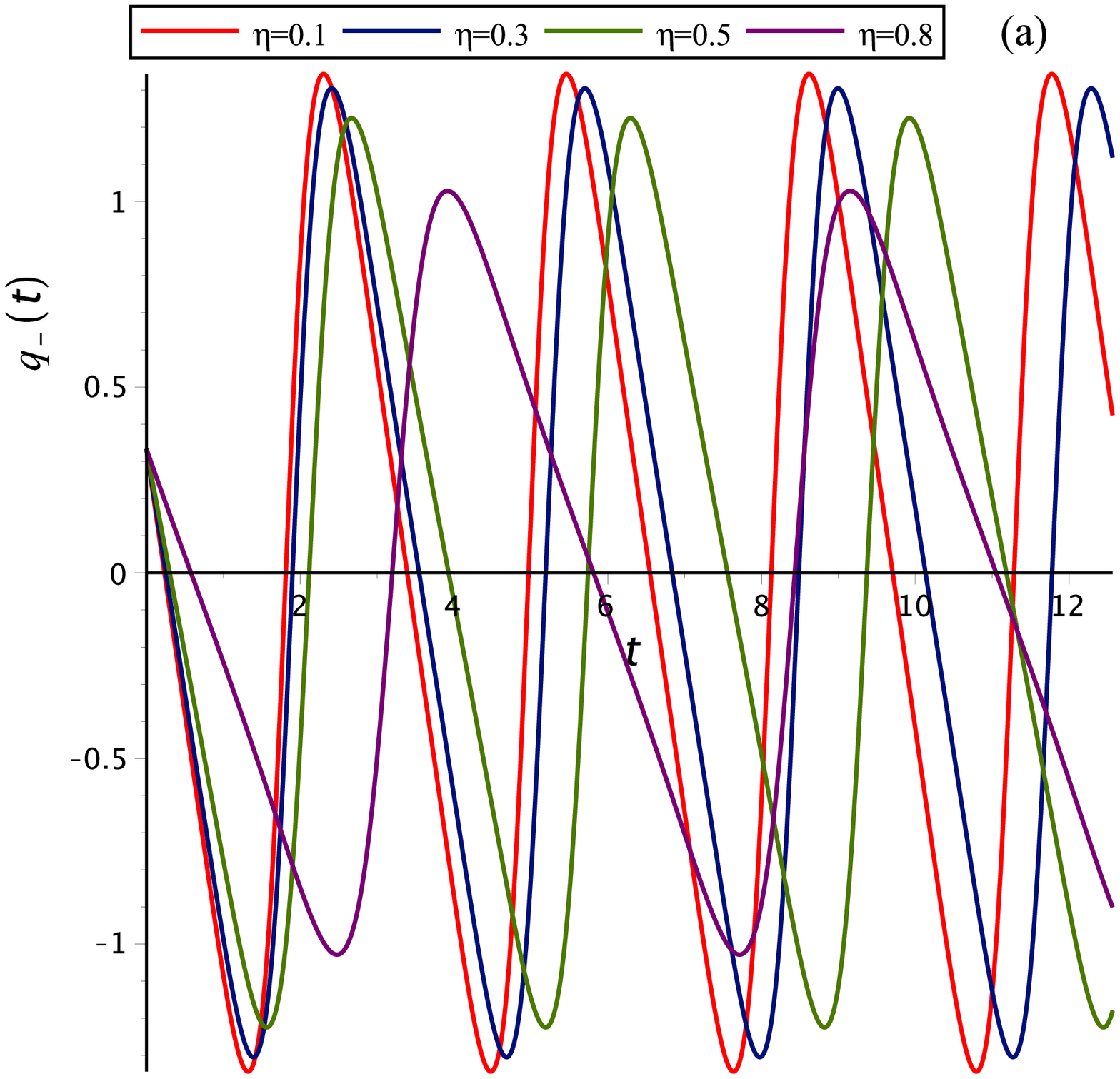}
\includegraphics[width=0.3\textwidth]{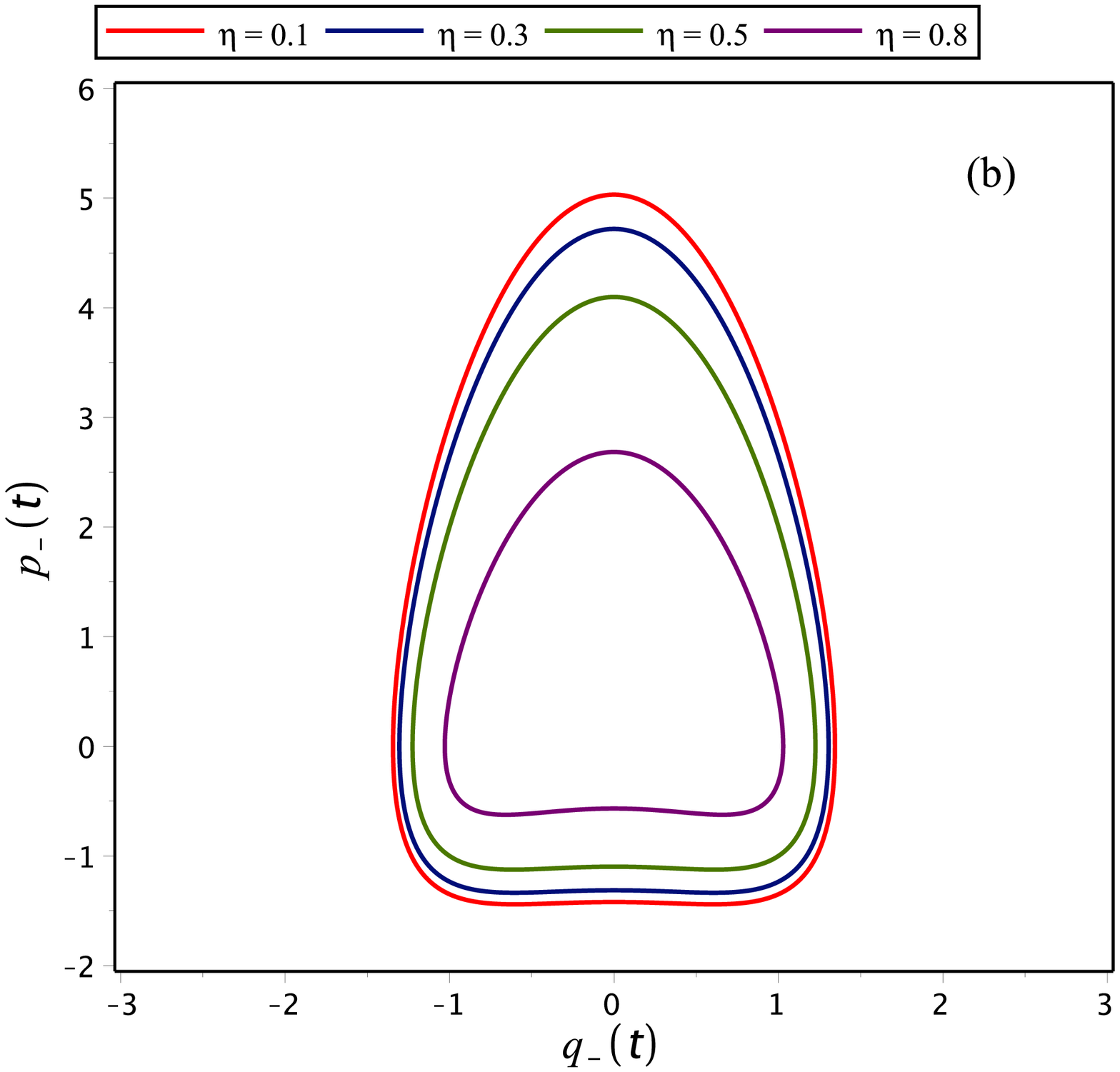} 
\includegraphics[width=0.3\textwidth]{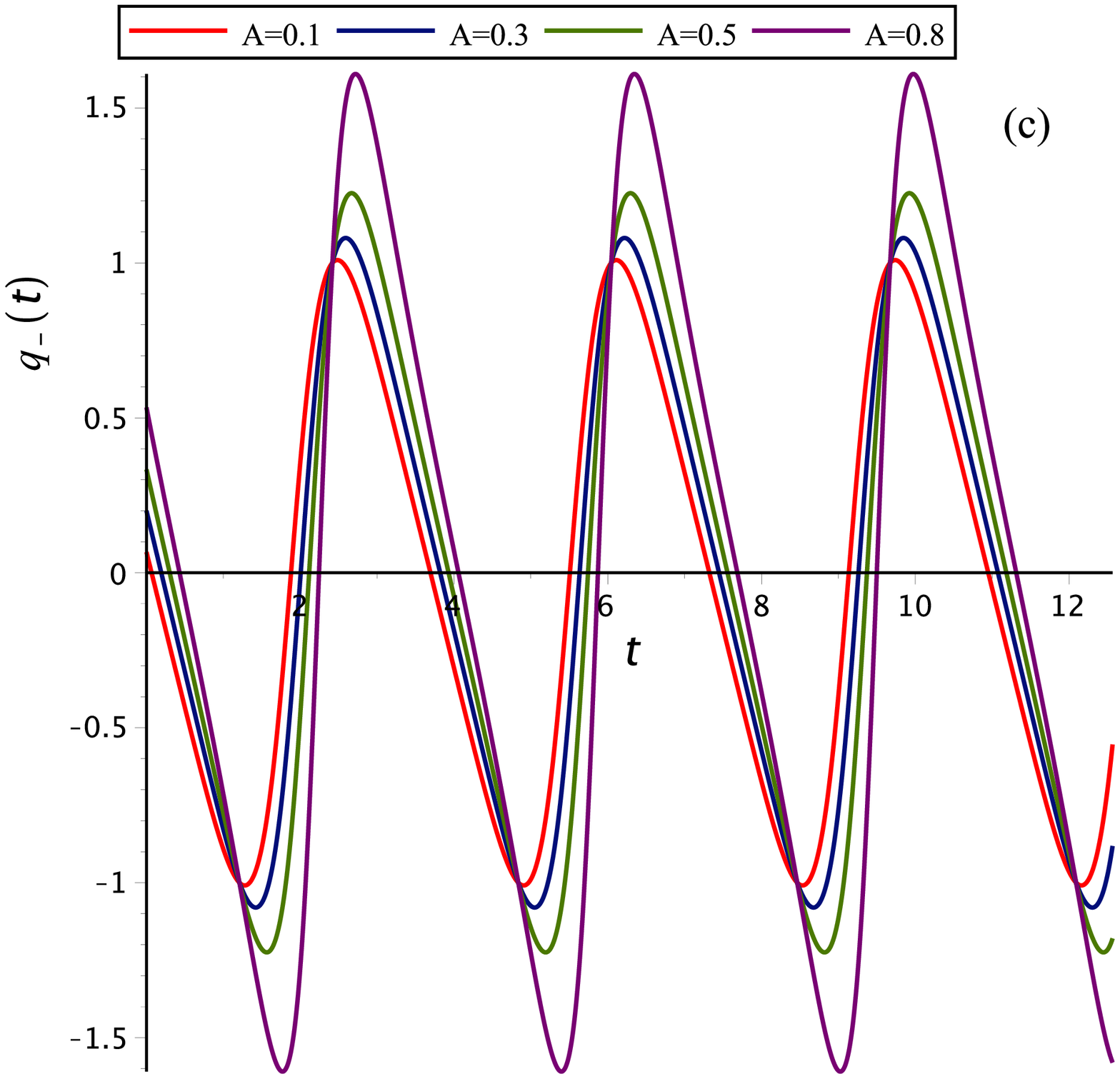}
\includegraphics[width=0.3\textwidth]{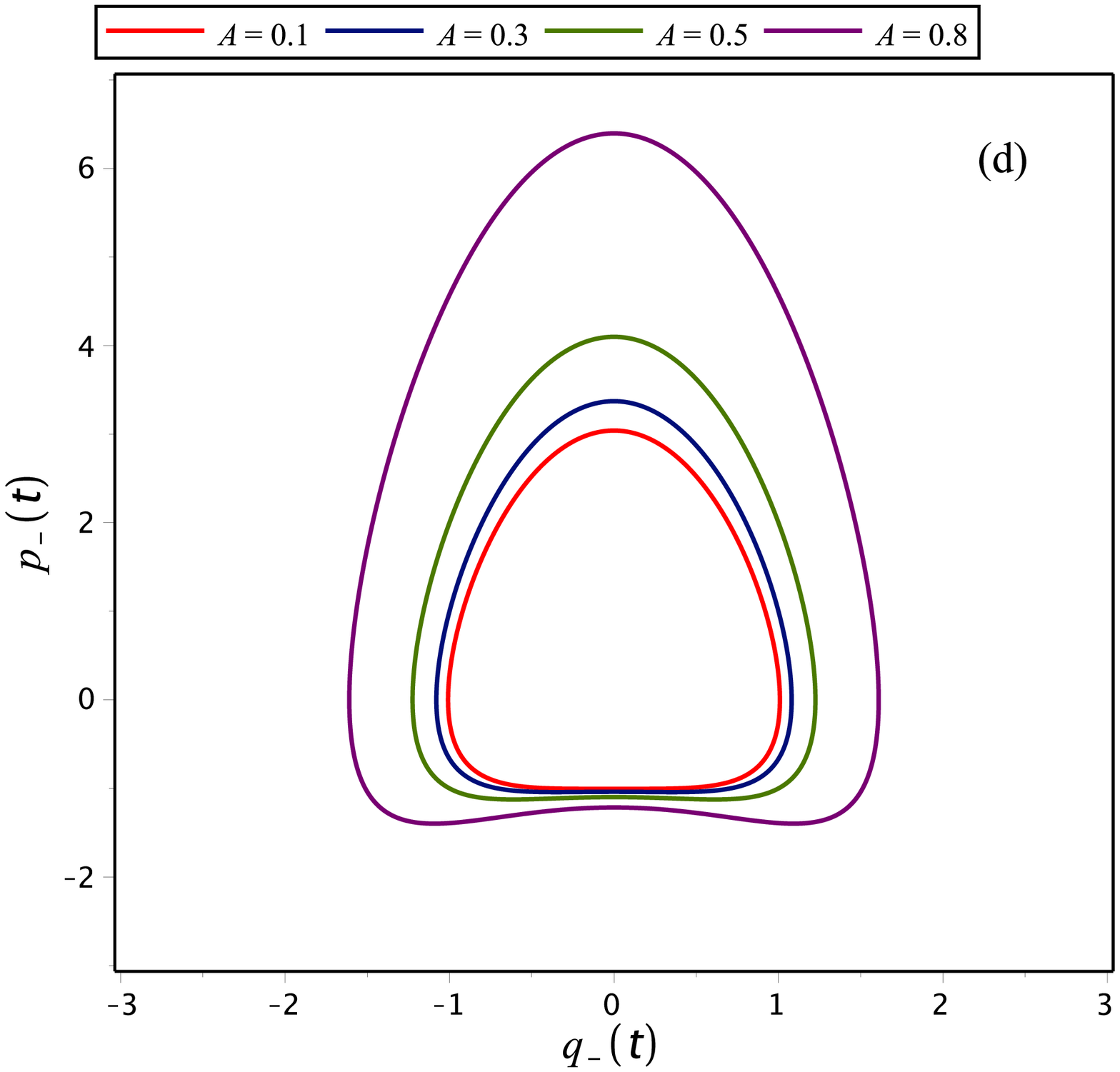}
\includegraphics[width=0.3\textwidth]{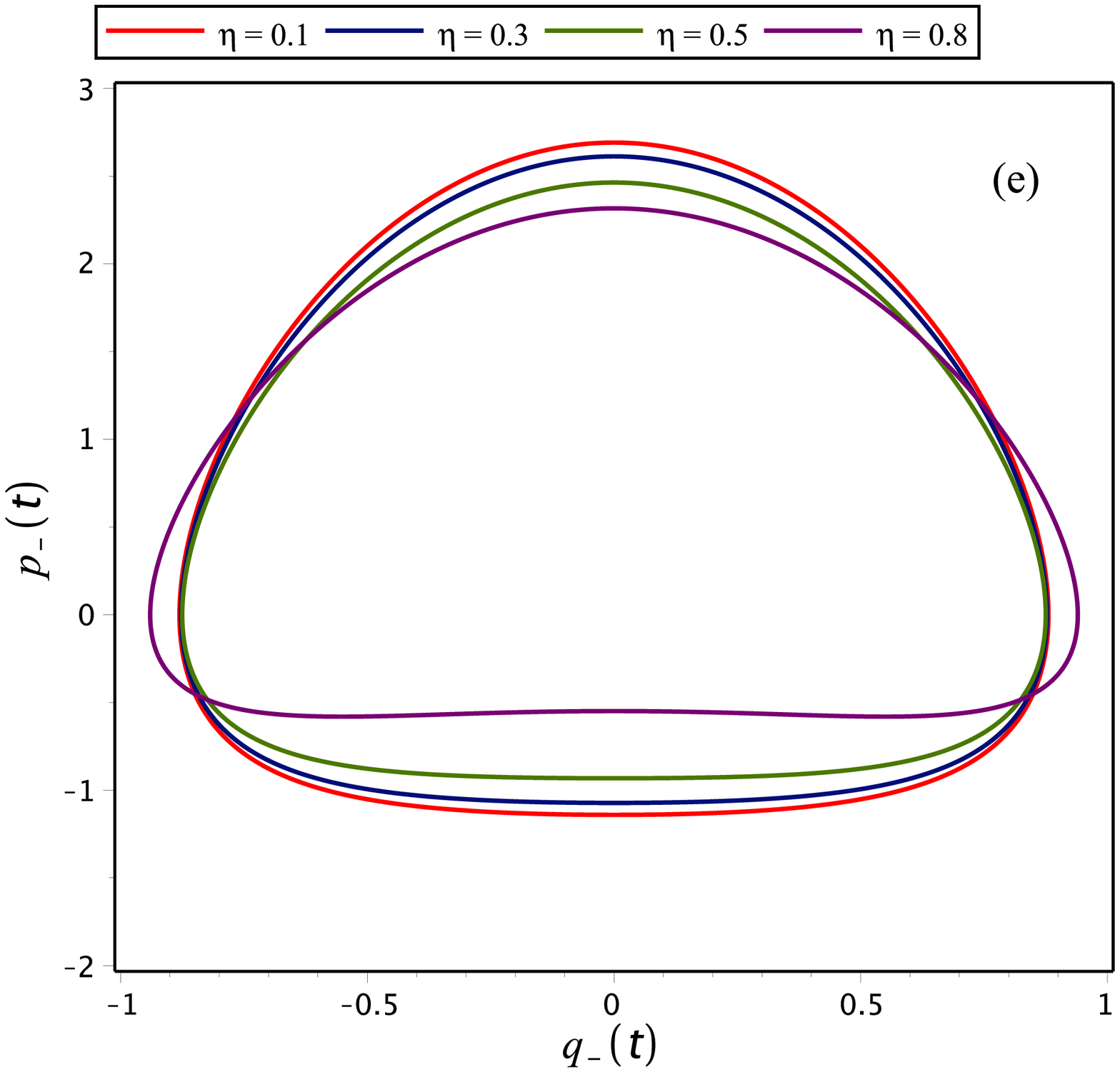}
\includegraphics[width=0.3\textwidth]{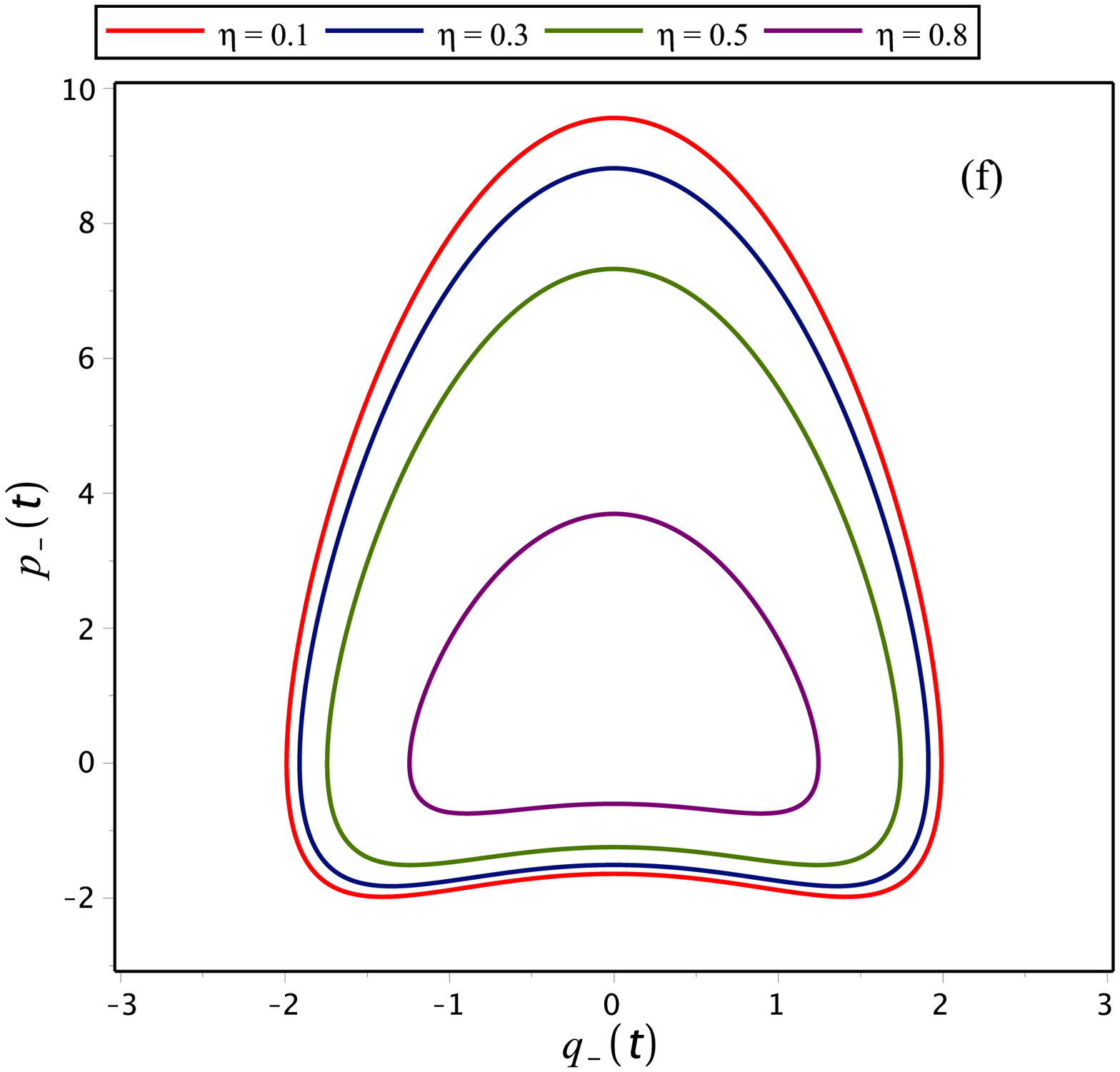}
\caption{\small 
{ For $A=B=0.5$, $\alpha=1$ and $\omega=2$ we show (a) $q_{_{-}}(t)$ of (\ref{MEE-Gsol}) for different $\eta<1$ values, and (b) phase-space trajectories $\{q_{_{-}}(t),p_{_{-}}(t)=\dot{q}_{_{-}}(t)\}$ of (\ref{MEE-Gsol}) for different values of $\eta<1$. For $B=0.5$, $\alpha=1$, $\omega=2$, and different values of $A$ we show (c)  $q_{_{-}}(t)$ of (\ref{MEE-Gsol}), and (d) phase-space trajectories  $\{q_{_{-}}(t),p_{_{-}}(t)=\dot{q}_{_{-}}(t)\}$ of (\ref{MEE-Gsol}). For $\alpha=1$, $\omega=2$ and $\eta<1$ we show (e) $\{q_{_{-}}(t),p_{_{-}}(t)=\dot{q}_{_{-}}(t)\}$ of (\ref{MEE-Gsol}) for $A=0.7>B=0.2$, and (f) $\{q_{_{-}}(t),p_{_{-}}(t)=\dot{q}_{_{-}}(t)\}$ of (\ref{MEE-Gsol}) for $B=0.7>A=0.2$.}}
\label{fig4}
\end{figure}%
\begin{equation}
U_{-}\left( t\right) =q_{_{-}}\left( t\right) \exp \left( \int \left( \alpha
q_{_{-}}\left( t\right) -\zeta \right) dt\right)   \label{U-MEE}
\end{equation}%
to obtain%
\begin{equation}
\ddot{q}_{_{-}}\left( t\right) +3\alpha q_{_{-}}\left( t\right) \dot{q}%
_{_{-}}\left( t\right) +\text{ }\tilde{\beta}^{2}\,q_{_{-}}\left( t\right)
+\alpha ^{2}q_{_{-}}\left( t\right) ^{3}=0\text{ };\text{ }\tilde{\beta}%
^{2}=\omega ^{2}-\zeta ^{2}.  \label{MEE1}
\end{equation}%
This nonlinear differential equation is the MEE and, herein, it describes a
classical particle, with mass $m_{\circ }=1$, moving under the influence of
the quartic anharmonic oscillator potential force field of (\ref{quartic
HO-potential}) and subjected to the Rayleigh dissipative and/or anti-dissipative force field%
\begin{equation}
\mathcal{R}_{_{-}}\left( q_{_{-}},\dot{q}_{_{-}}\right) =\frac{3}{2}\alpha
q_{_{-}}\left( t\right) \;\dot{q}_{_{-}}\left( t\right) ^{2}-\zeta
^{2}q_{_{-}}\left( t\right) \dot{q}_{_{-}}\left( t\right) .  \label{R3}
\end{equation}%
Then the standard Lagrangian for such a system is readily given in (\ref{GMEE-L1}) and the corresponding nonlinear dynamical equation (\ref{GMEE2}) is obtained through the Euler-Lagrange recipe of (\ref{Euler-Lagrange}).
Nevertheless, the nonlinear dynamical equation (\ref{MEE1}) admits an exact
solution in the form of%
\begin{equation}
q_{_{-}}\left( t\right) =\frac{\tilde{\beta}}{\alpha }\left\{ \frac{%
C_{2}\cos \left( \tilde{\beta}t\right) -\sin \left( \tilde{\beta}t\right) }{%
C_{1}+C_{2}\sin \left( \tilde{\beta}t\right) +\cos \left( \tilde{\beta}%
t\right) }\right\} \text{ };\tilde{\beta}=i\beta =i\omega \sqrt{\eta ^{2}-1}.
\label{MEE-sol}
\end{equation}%
With $B=1/C_{1}$, $A=\tilde{\beta}C_{2}/C_{1}$, $\zeta =\omega \eta$ and $%
\tilde{\beta}=i\beta$ one obtains%
\begin{equation}
q_{_{-}}\left( t\right) =\frac{1}{\alpha }\left\{ \frac{A\cos \left( \tilde{%
\beta}t\right) -B\tilde{\beta}\sin \left( \tilde{\beta}t\right) }{1+\frac{A}{%
\tilde{\beta}}\sin \left( \tilde{\beta}t\right) +B\cos \left( \tilde{\beta}%
t\right) }\right\} \Longleftrightarrow q_{_{-}}\left( t\right) =\frac{1}{%
\alpha }\left\{ \frac{A\cosh \left( \beta t\right) +B\beta \sinh \left(
\beta t\right) }{1+\frac{A}{\beta }\sinh \left( \beta t\right) +B\cosh
\left( \beta t\right) }\right\}   \label{MEE-Gsol}
\end{equation}%
This result when substituted in (\ref{U-MEE}) would yield%
\begin{equation}
U_{-}\left( t\right) =\frac{1}{\alpha }\left[ A\cosh \left( \beta t\right)
+B\beta \sinh \left( \beta t\right) \right] e^{-\omega \eta t}
\label{DHO-Gsol}
\end{equation}%
as the exact solution of (\ref{DHO-eq}). Obviously, the solutions (\ref{MEE-Gsol}) as well as (\ref{DHO-Gsol}) suggest that the parameter $\alpha $ , in this case, may take any value provided that $\alpha \in 
\mathbb{R}
$.  Moreover, we observe that the interesting solution reported by Chandrasekar et al. \cite{Chandrasekar 2012} (obtained by using $U(t)=A\, sin(\omega t)$ as the solution for the HO and mapped it into MEE) is a especial case of our general solution in (\ref{MEE-Gsol}), where our $\tilde{\beta}=\omega$ of  \cite{Chandrasekar 2012}.

In Figure 4, for  $\omega=2$, $\alpha=1$, $A=B=0.5$, and different values of $\eta<1$,  we plot (a) $q_{_{-}}(t)$ of (\ref{MEE-Gsol}), and (b) phase-space trajectories $\{q_{_{-}}(t),p_{_{-}}(t)=\dot{q}_{_{-}}(t)\}$ of (\ref{MEE-Gsol}).  For $B=0.5$, $\alpha=1$, $\omega=2$, and different values of $A$ we show (c)  $q_{_{-}}(t)$ of (\ref{MEE-Gsol}), and (d) phase-space trajectories  $\{q_{_{-}}(t),p_{_{-}}(t)=\dot{q}_{_{-}}(t)\}$ of (\ref{MEE-Gsol}). For $\alpha=1$, $\omega=2$ and $\eta<1$ we show (e) $\{q_{_{-}}(t),p_{_{-}}(t)=\dot{q}_{_{-}}(t)\}$ of (\ref{MEE-Gsol}) for $A=0.7>B=0.2$, and (f) $\{q_{_{-}}(t),p_{_{-}}(t)=\dot{q}_{_{-}}(t)\}$ of (\ref{MEE-Gsol}) for $B=0.7>A=0.2$. Obviously, the oscillations are non-isochronic as documented in (a) and (c). Moreover, whilst the phase-space trajectories in (b), (d), and (f) replicate those of the MEE  \cite{Chandrasekar 2012}, the phase-space trajectories in (e) exhibit classical states  $\{q_{_{-}}(t),p_{_{-}}(t)\}$  crossings for $A=0.7>B=0.2$.  No classical states crossings for  $B\leq A$ are observed (as documented in (b), (d), and (f)). It is, therefore, necessary and sufficient to use the most general solution (\ref{MEE-Gsol}) for MEE so that one carries out comprehensive analysis of such a commonly used  dynamical system  (note that our $\tilde{\beta}$ corresponds to $\omega$ in \cite{Chandrasekar 2012}).

\section{Concluding remarks}

In the current methodical proposal, we have considered the linearization of some GMEE-type (\ref{GMEE0}) dynamical equations into HO and DHO type ones and reported their exact solutions. We have shown that structure of the nonlocal transformation and the linearizability into HO or DHO determine the nature/structure of the dynamical forces involved (consequently determine the form of the dynamical equation). The details of the obtained solutions suggest that they are explicit,  general and valid solutions not only for the GMEE-type (i.e., $q_{_{\pm}}(t)$)  equations (\ref{GMEE01}) but also for the HO and DHO type (i.e., $U_{_{\pm}}(t)$)  ones. Moreover, we have considered the linearization of a MEE-type (\ref{GMEE2}) dynamical equations into DHO-type ones and reported their exact solutions. Therein, the reported solutions turned out to be explicit, general and valid for the MEE-type (i.e., $q_{_{-}}(t)$) and for the DHO-type (i.e., $U_{_{-}}(t)$) equations. Yet, for each dynamical system we have reported illustrative figures for each  $q_{_{\pm}}(t)$ and $\{ q_{_{\pm}}(t), p_{_{\pm}}(t)\}$ classical state (i.e., phase-space trajectory) as they evolve in time. To the best of our knowledge, the reported solutions, in both section 2 and 3, have never been reported elsewhere.%
\begin{figure}[!ht]  
\centering
\includegraphics[width=0.2\textwidth]{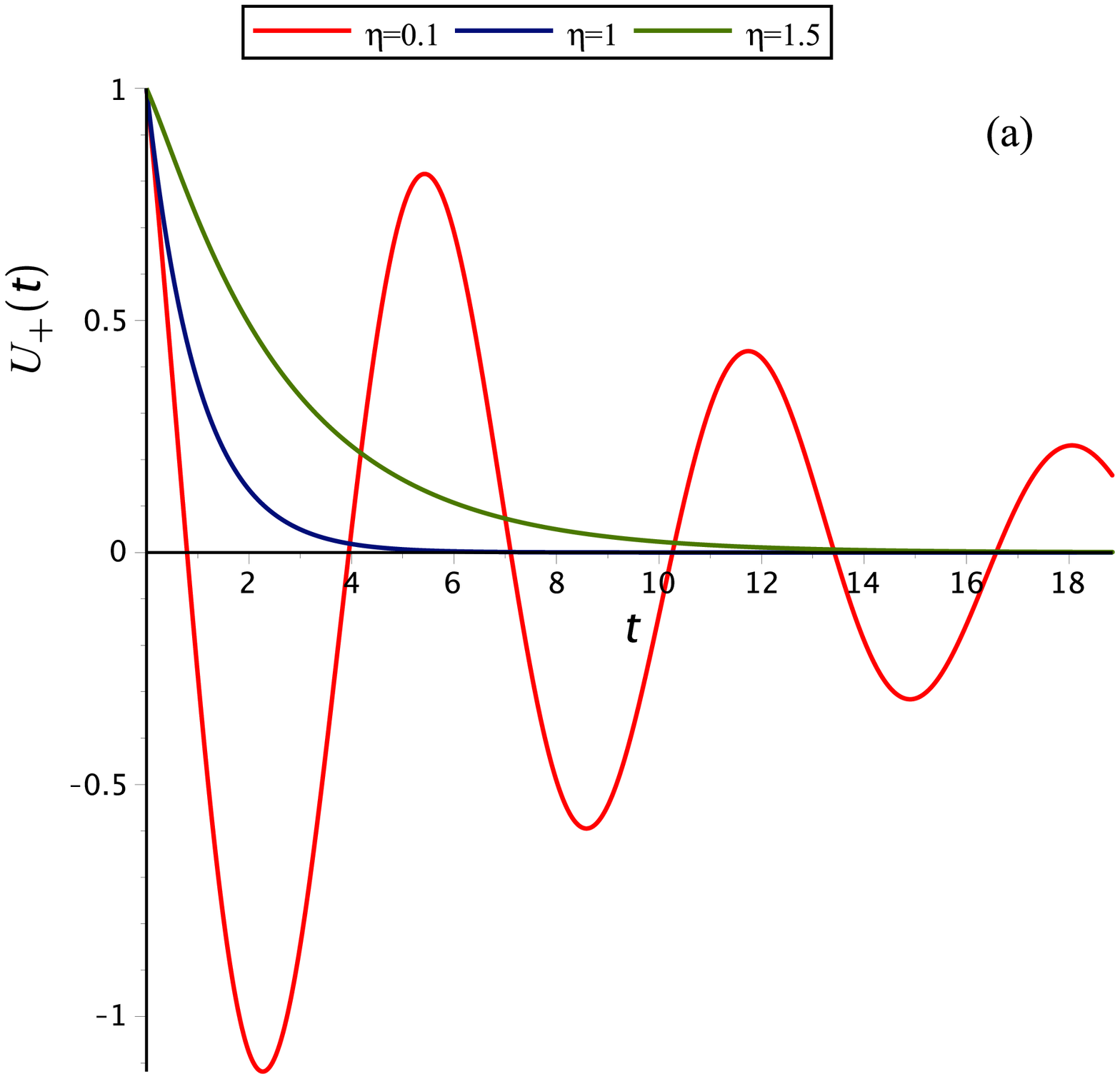}
\includegraphics[width=0.2\textwidth]{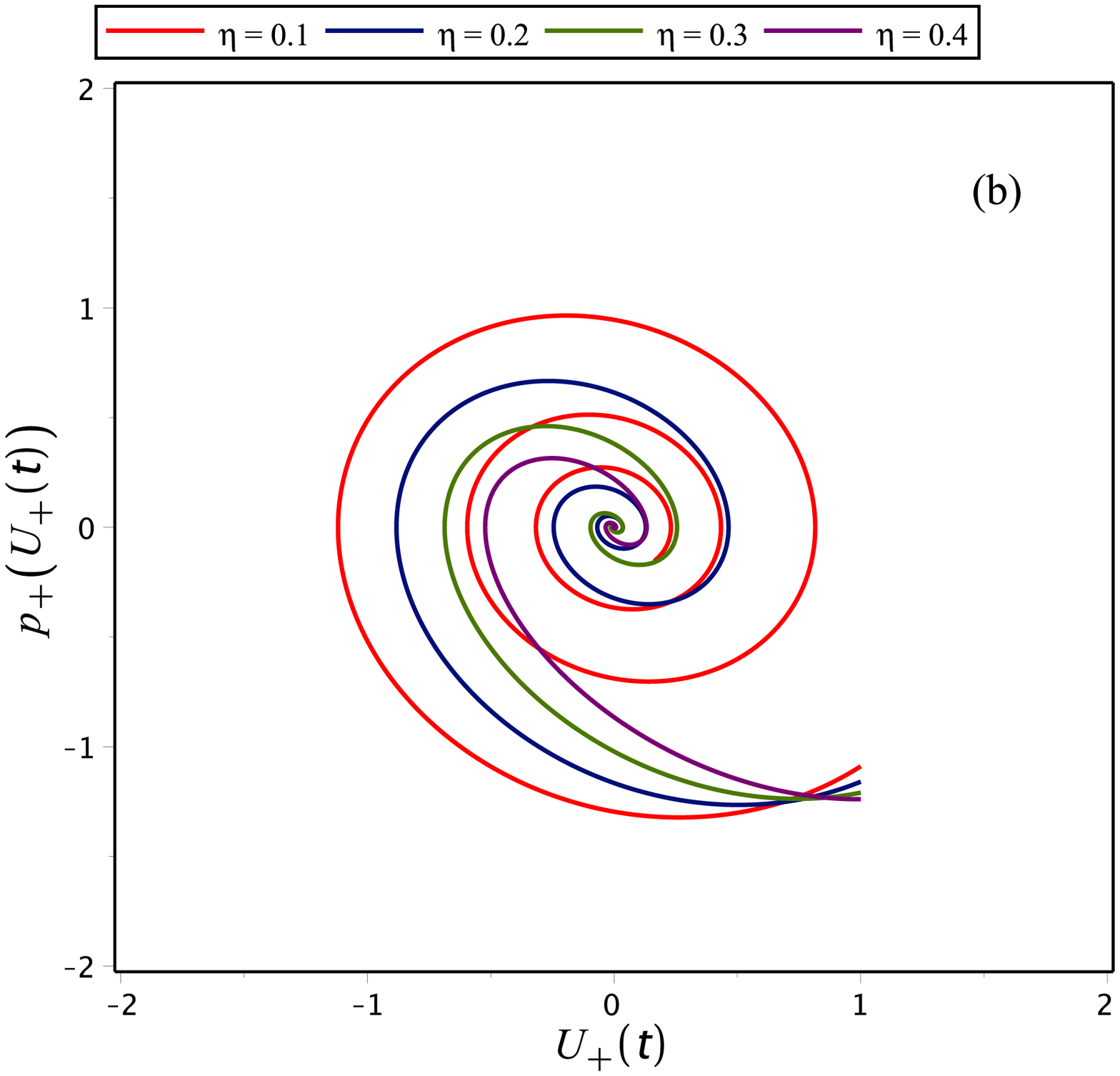}
\includegraphics[width=0.2\textwidth]{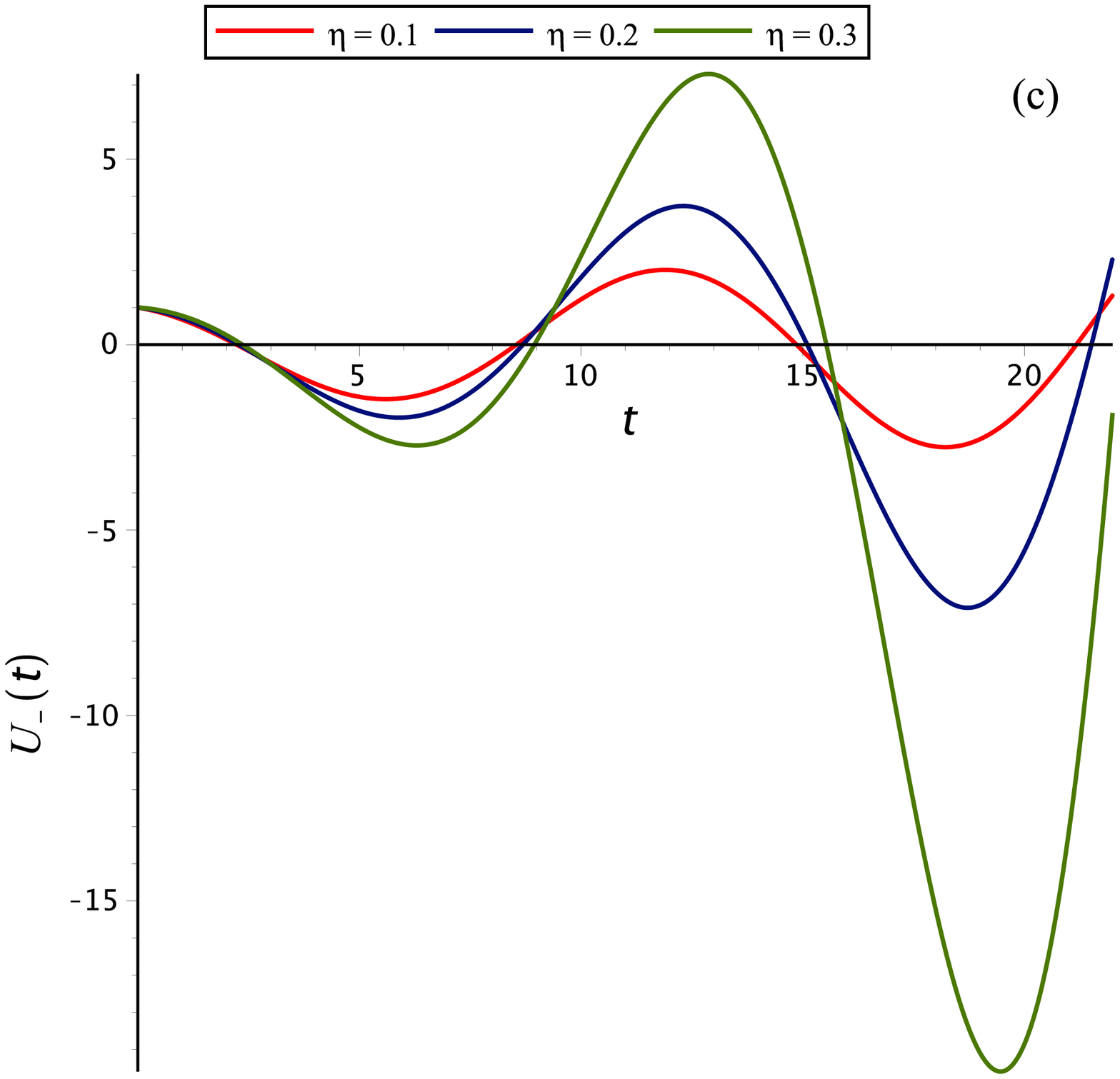}
\includegraphics[width=0.2\textwidth]{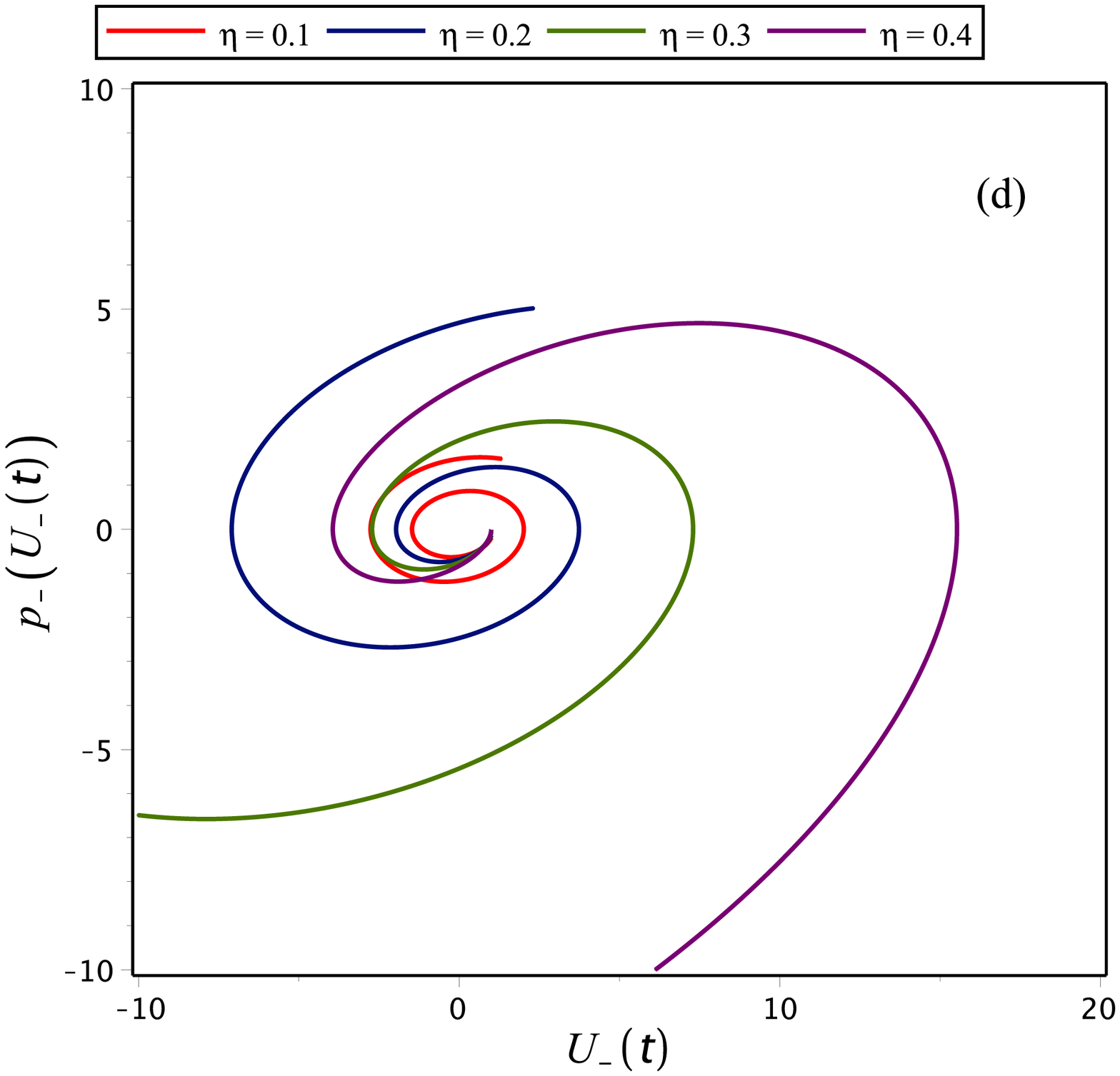}
\caption{\small 
{ For the DHO-equation in (\ref{DHO1-eq}) with $A=B=1$ and $\omega=1$, we show (a) $U_{_{+}}(t)$ of (\ref{U1-DHO-Gsol}) for $\eta=0.1$ (under-damping), $\eta=1$ (critical-damping), and for $\eta=1.5$ (over-damping), and  (b) the phase-space trajectories $\{U_{_{+}}(t),p_{_{+}}(t)=\dot{U}_{_{+}}(t)\}$ for different values of $\eta<1$. For $A=B=1$ and $\omega=0.5$, we show (c) $U_{_{-}}(t)$ of (\ref{U1-DHO-Gsol}) for different values of $\eta<1$, and (d) the phase-space trajectories $\{U_{_{+}}(t),p_{_{-}}(U_{_{-}}(t))=\dot{U}_{_{-}}(t)\}$ for different values of $\eta<1$. }}
\label{fig5}
\end{figure}%

In the linearization process, nevertheless, and as a byproduct of our reverse engineering strategy, we have started from the general solutions for the GMEE (\ref{GMEE1-sol}) and (\ref{GMEE2-sol}) to come out with the most general solutions for the HO (\ref{HO-sol}), DHO (\ref{DHO-sol}) and (\ref{DHO-Gsol}). The solution  $U_{_{\pm }}\left(
t\right) $ in (\ref{HO-sol})  for the HO in (\ref{HO-eq}) turned out to be a textbook one, provided that $a=C\eta $ and $b=\pm C$., that provides a consistent nonlocal connection, through the nonlocal transformation in (\ref{NL1}), to the generalized modified Emden equation (\ref{GMEE1}) along with its exact general solution (\ref{GMEE1-sol}). However, for the DHO one may eventually absorb the parameters in the related amplitudes of $U_{+}\left( t\right) $ in (\ref{DHO-sol}) and of $U_{-}\left(
t\right) $ in (\ref{DHO-Gsol}) and cast the most general solution for the DHO problem (\ref{DHO-eq}) as%
\begin{equation}
U\left( t\right) =\left[ A\cosh \left( \beta t\right) +B%
\beta \sinh \left( \beta t\right) \right] e^{-\omega \eta t}\text{ };  \beta =\omega \sqrt{\eta ^{2}-1},  \label{U-DHO-Gsol}
\end{equation}%
where we have used%
\begin{eqnarray}
A=\left\{ 
\begin{tabular}{ll}
$2A$ &; of $U_{+}\left( t\right) $ in (\ref{DHO-sol}) \\ 
$\frac{A}{\alpha}$ &; of $U_{-}\left( t\right) $ of (\ref{DHO-Gsol})  
\end{tabular}%
\right\},
\label{At}
\end{eqnarray}%
\begin{eqnarray}
B=\left\{ 
\begin{tabular}{ll}
$-\frac{A}{\omega \eta}$ &; of $U_{_{+}}\left( t\right)$  in (\ref{DHO-sol}) \\ 
$\frac{B}{\alpha}$ &; of $U_{_{-}}\left( t\right)$  of (\ref{DHO-Gsol}) 
\end{tabular}%
\right\}. 
\label{Bt}
\end{eqnarray}%
Hence, one may now use $\eta <1$ (for under damping), $\eta =1$ (for critical damping), and $\eta >1$ (for over damping) without any consequential complex settings in $U(t)$ and/or in $\dot{U}(t)$. This would also allow us to choose proper initial conditions on  $U(0)$ and/or $\dot{U}(0)$ associated with the DHO problem at hand. 

Yet, one may rewrite the DHO equation (\ref{DHO-eq}) to include both damped and anti-damped harmonic oscillators as
\begin{equation}
\ddot{U}_{_{\pm}}\left( t\right) \pm2\zeta \dot{U}_{_{\pm}}\left( t\right) +\omega ^{2}U_{_{\pm}}\left(
t\right) =0,  \label{DHO1-eq}
\end{equation}%
then the general solution in (\ref{U-DHO-Gsol}) would read%
\begin{equation}
U_{_{\pm}}\left( t\right) =\left[ A\cosh \left( \beta t\right) +B%
\beta \sinh \left( \beta t\right) \right] e^{\mp\omega \eta t}\text{ }.  \label{U1-DHO-Gsol}
\end{equation}%
Such damped or anti-damped linear harmonic oscillator systems are eminent (also physically viable) consequences of the linearization process discussed above. In Figure 5, we show the behaviour of such general solution (\ref{U1-DHO-Gsol}).  For $A=B=1$ and $\omega=1$, we show (a) $U_{_{+}}(t)$ of (\ref{U1-DHO-Gsol}) for $\eta=0.1$ (under-damping), $\eta=1$ (critical-damping), for $\eta=1.5$ (over-damping), and  (b) the phase-space trajectories $\{U_{_{+}}(t),p_{_{+}}(U_{_{+}}(t))=\dot{U}_{_{+}}(t)\}$ for different values of $\eta<1$. For $A=B=1$ and $\omega=0.5$, we show (c) $U_{_{-}}(t)$ of (\ref{U1-DHO-Gsol}) for different values of $\eta<1$, and (d) the phase-space trajectories $\{U_{_{+}}(t),p_{_{-}}(U_{_{-}}(t))=\dot{U}_{_{-}}(t)\}$ for different values of $\eta<1$. Obviously, the typical behaviour of the DHO and anti-DHO are exhibited therein.

Finally, it would be interesting if further coordinate transformation (e.g., $q_{_{\pm}}(t)\rightarrow \sqrt{Q(x(t))}x(t)$) is carried out to effectively yield position-dependent mass (PDM) settings (c.f., e.g.,  \cite{Mustafa PS 2021,Mustafa 2015,Mustafa Algadhi 2019,Nabulsi1 2020,Nabulsi2 2020,Quesne 2019,da Costa1 2020,Ranada 2016,Carinena Herranz 2017}). One would then come out with a mixed type Li\'{e}nard equation to be mapped into GMEE type (which is effectively a linear type  Li\'{e}nard equation) and consequently into HO or DHO.  The new PDM system would then inherit the exact solutions reported above.


\end{document}